\documentclass[lettersize,journal]{IEEEtran}
\usepackage[T1]{fontenc}% optional T1 font encoding

\usepackage{amsmath}
\usepackage[cmintegrals]{newtxmath}
\usepackage{bm}

\newtheorem{Remark}{\it Remark}[section]

\newtheorem{Proposition}{\it Proposition}[section]
\newtheorem{Lemma}{\it Lemma}[section]

\makeatletter
\newcommand{\Rmnum}[1]{\expandafter\@slowromancap\romannumeral #1@}
\makeatother
\usepackage{verbatim}
\usepackage{graphicx}
\usepackage{multicol}
\usepackage{setspace}
\usepackage{amsmath}
\usepackage{epstopdf}
\usepackage{fancyhdr} 
\usepackage{enumerate}

\usepackage{amssymb}
\usepackage{stfloats}
\usepackage{multirow}
\usepackage{threeparttable}
\usepackage[justification=centering]{caption}
\usepackage{makecell}
\usepackage{caption}
\usepackage{cases}
\usepackage{algorithm}
\usepackage{algpseudocode}
\usepackage{graphics}
\usepackage{float}
\usepackage{subfig}
\usepackage{epsfig}
\usepackage{float}
\usepackage{color}
\usepackage{colortbl}
\usepackage{cite}

\begin{document}
	
	\title{A Joint Model and Data Driven Method for Distributed Estimation}
	\author{Meng~He,~\IEEEmembership{Member,~IEEE},
	Ran~Li,~\IEEEmembership{Member,~IEEE},
	Chuan~Huang,~\IEEEmembership{Member,~IEEE},
	and Shulong Zhang%
		\thanks{
			Part of this paper was presented in IEEE International Conference on Communications in China (ICCC) 2022\cite{9880805}. This work was supported in part by the Natural Science Foundation of China under Grant No. 62022070 and No. 62341112, in part by Shenzhen high-tech zone project No. KC2022KCCX0041, in part by the key project of Shenzhen No. JCYJ20220818103006013, in part by the Shenzhen Outstanding Talents Training Fund 202002,  in part by the Guangdong Provincial Key Laboratory of Future Networks of Intelligence (Grant No. 2022B1212010001), and in part by the Shenzhen Key Laboratory of Big Data and Artificial Intelligence (Grant No. ZDSYS201707251409055). (Corresponding author: Chuan Huang).
			
			M. He, R. Li and C. Huang are with the Future Network of Intelligence Institute and the School of Science and Engineering, The Chinese University of Hong Kong, Shenzhen, China, 518172. Chuan Huang is also with Peng Cheng Laboratory, Shenzhen, China, 518066. (Emails: menghe@link.cuhk.edu.cn, ranli2@link.cuhk.edu.cn, and huangchuan@cuhk.edu.cn).
			
			S. Zhang is with the SF Technology, Shenzhen, China, 518052. (Email: shulongzhang@sf-express.com).
		}
	}% <-this % stops a space
		
%	\author{
%	\IEEEauthorblockN{
%	Meng~He\IEEEauthorrefmark{1}, 
%	Chuan~Huang\IEEEauthorrefmark{1}\IEEEauthorrefmark{2}, 
%	and Shengpei~Jiang\IEEEauthorrefmark{3}\\
%	\IEEEauthorblockA{\IEEEauthorrefmark{1}School of Science and Engineering (SSE) and Future Network of Intelligence Institute (FNii),\\ 
%	The Chinese University of Hong Kong, Shenzhen 518172, China}
%	\IEEEauthorblockA{\IEEEauthorrefmark{2}Peng Cheng Laboratory, Shenzhen 518066, China}
%	\IEEEauthorblockA{\IEEEauthorrefmark{3}SF Technology, Shenzhen 518052, China}
%	Emails: \IEEEauthorrefmark{1}menghe@link.cuhk.edu.cn, \IEEEauthorrefmark{1}\IEEEauthorrefmark{2}huangchuan@cuhk.edu.cn, \IEEEauthorrefmark{3}philip.jiang@sfmail.sf-express.com}
%	}%
	
	\maketitle
	\begin{abstract}
	This paper considers the problem of distributed estimation in wireless sensor networks (WSN), which is anticipated to support a wide range of applications such as the environmental monitoring, weather forecasting, and location estimation. To this end, we propose a joint model and data driven distributed estimation method by designing the optimal quantizers and fusion center (FC) based on the Bayesian and minimum mean square error (MMSE) criterions.	First, universal mean square error (MSE) lower bound for the quantization-based distributed estimation is derived and adopted as the design metric for the quantizers. Then, the optimality of the mean-fusion operation for the FC with MMSE criterion is proved. Next, by exploiting different levels of the statistic information of the desired parameter and observation noise, a joint model and data driven method is proposed to train parts of the quantizer and FC modules as deep neural networks (DNNs), and two loss functions derived from the MMSE criterion are adopted for the sequential training scheme. Furthermore, we extend the above results to the case with multi-bit quantizers, considering both the parallel and one-hot quantization schemes. Finally, simulation results reveal that the proposed method outperforms the state-of-the-art schemes in typical scenarios.
	\end{abstract}


	\IEEEpeerreviewmaketitle
	
	\section{Introduction}
	Motivated by the high-speed development of the wireless sensor network (WSN) in practical applications \cite{kottas2012spatial,french2013spatio}, such as environmental monitoring, weather forecasts, and health care, the area of parameter estimation from distributed data has been thoroughly investigated in bunches of works\cite{5280228,4407653,4490095,1597575}. A typical distributed estimation network consists of multiple sensors deployed at different locations, each observing the desired unknown parameter and transmitting the local observations to the fusion center (FC)\cite{1619423}. Constrained by the limited wireless communication resources \cite{5494187,7450677}, e.g., bandwidth and energy, the local observations at the sensors are usually quantized into finite number of bits before being transmitted to the FC\cite{1657815,7111345,7815400}. The FC leverages the quantized information received from all the sensors to estimate the desired parameter. 
	
	The performance of quantization-based distributed estimation methods and the optimization of distributed estimation network have been extensively investigated in the literatures \cite{241739,243470,1643916,1435664,4156398}.
	Considering the impact of transmission errors, the authors in \cite{ghazanfari2015formulation} formulated the distributed estimation problem in the context of least mean square adaptive networks and derived the closed-form expressions for the steady-state mean square error (MSE) of the network.
	Considering the different distortion criteria, e.g., Fisher information \cite{21219} and minimum mean square error (MMSE) \cite{schonhoff2006detection}, the authors in \cite{241739} investigated the optimal quantizer design for decentralized parameter estimation with two distributed sensors.
	Based on the generalization of the classical Lloyd-Max results \cite{1056489,1057548}, the author in \cite{243470} proposed an algorithm to design the optimal non-linear distributed estimators with bivariate probability distribution. Considering the constraint of one-bit quantization at the sensors, the authors in \cite{1643916} established the asymptotic optimality of the maximum-likelihood (ML) estimator for the deterministic mean-location parameter estimation problem. 
	However, the above works were limited to the assumption of either a deterministic desired parameter or a random one with perfect information of its distribution.
	
	Based on the minimization of Cramer-Rao lower bound (CRLB), the probabilistic quantization strategies have been actively investigated recently \cite{7094709,8334277,5184907,6174480,6882252,7472370}. Considering the one-bit quantization scheme at the sensors and applying the minimax criterion, the authors in \cite{5184907} derived the minimax CRLB for the scenario with ideal noiseless observations. Following the identical one-bit quantization scheme across different sensors, the authors in \cite{6174480} further approximated the optimal probabilistic quantization by a parameterized antisymmetric and piecewise-linear function to minimize the corresponding minimax CRLB. 
	In \cite{6882252}, the authors obtained the optimality conditions for using binary quantizer at all sensors and the optimal binary quantizer for the location parameter estimation problem. 
	However, these works were limited to the noiseless observation scenario and the extension to the noisy case was subject to the assumption of perfect knowledge about the distributions of the desired parameter and observation noise.
	The optimal quantizer for noisy scenario with imperfect statistic information remains to be further explored.
	
	This paper considers a WSN in which local observations are acquired at distributed sensors and transmitted to the FC for the estimation of a desired parameter. Due to the bandwidth constraints at the sensors, these observations are quantized into finite bits prior to transmissions. The FC aims to estimate the desired parameter based on the quantized information from all sensors. The major contributions of this paper are summarized as follows:
	\begin{itemize}
		\item First, considering the one-bit quantization constraint at sensors, we derive the universal estimation MSE lower bound for the case with conditionally independent and identically distributed (i.i.d.) observations. This lower bound is shown to be independent of the FC design, and thus the binary probabilistic quantizer is designed to minimize this MSE lower bound.
		
		\item Second, we prove the optimality of the mean-fusion operation at the FC, which uses only the average of all the quantized data for estimation, in terms of achieving the same MSE lower bound as the one by using all the quantized data from the sensors. Thus, a FC module with mean-fusion operation on the quantized data is proposed to accommodate the system with a variable number of sensors.
		
		\item 
		Then, considering the scenarios that the distributions of both the desired parameter and noise are unknown or only the noise distribution is known, a joint model and data driven method is proposed to train the probability controller in quantizer and the estimator in FC as  deep neural networks (DNNs). Two loss functions derived from MMSE criterion are utilized for the sequential training of their design parameters.
		
		\item 
		Finally, we extend the above results to the case with multi-bit quantizer. We propose the multi-bit parallel and one-hot quantization schemes, and analyze the corresponding MSE lower bounds and optimality of mean-fusion operations, respectively.
	\end{itemize}
	
	The remainder of the paper is organized as follows: Section II introduces the system model and presents the problem formulation. Section III proposes the joint model and data driven method for binary quantization. Section IV extends the results to the scenario of multi-bit quantization. Section V shows the simulation analysis. Finally, section VI concludes this paper.
	
	Notations: 
	Boldface lowercase and uppercase letters, e.g., $ \mathbf{x}$ and $\mathbf{X}$, denote vectors and matrix, respectively. $[\mathbf{X}]_{i,j}$ denotes the $(i,j)$-th entry of the matrix $X$.  $\mathbb{E}[\cdot]$ represents expectation. $|x|$ denotes the absolute value of scalar $x$. $\mathbb{N}^+$ is the positive integer set. $[N]$ denotes the set of positive integers no bigger than $N$. Copperplate uppercase letters, e.g., $\mathcal{U}$, denotes the set. $U_1 \circ U_2$ represents the composition of neural networks $U_1$ and $U_2$.
	
	\section{System Model and Problem Formulation}
	\label{Problem Formulation}	
	\subsection{System model}
\begin{figure}[htbp]
		\vspace{0pt}
		\setlength{\abovecaptionskip}{10pt}
		\setlength{\belowcaptionskip}{0pt}
		\centering
		\includegraphics[scale=0.8]{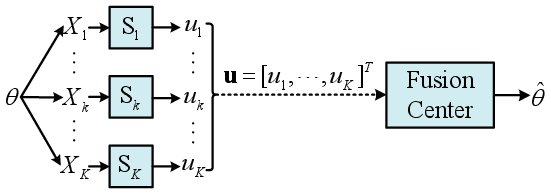}
		\caption{System Model for Distributed Estimation.}
		\label{System Model}
	\end{figure}

	A generalized distributed estimation problem is considered in this paper as shown in Fig. \ref{System Model}, where the FC aims to estimate the desired parameter $\theta$ by using $K$ distributed sensors, denoted as $\text{S}_1,\cdots,\text{S}_K$. Sensor $\text{S}_k$, $k=1,2,\cdots,K$, observes $\theta$ independently and obtains its local observation $X_k$, which is a noisy version of $\theta$. 
	Here, we consider a widely adopted scenario that the observation noises at all sensors are i.i.d., and thus the local observations obtained at all sensors are conditionally i.i.d. with given $\theta$, i.e.,
	\begin{equation}
	 \label{eq: conditionally i.i.d. local observations 1}
	 f_{X_1,\cdots,X_K}(x_1,\cdots,x_K|\theta)=\prod_{k=1}^{K}f_{X_k}(x_k|\theta)=\prod_{k=1}^{K}f_{X}(x_k|\theta),
	\end{equation}
	$\forall x_1,\cdots,x_K\in\mathbb{R}$, where $f_{X_1,\cdots,X_K}(\cdot|\theta)$ is the conditional joint probability density function (PDF) of all observations with given $\theta$ and $f_X(\cdot|\theta)=f_{X_1}(\cdot|\theta)=\cdots=f_{X_K}(\cdot|\theta)$ is the conditional marginal PDF of the local observation with given $\theta$ at any sensor. 
	
	Due to the bandwidth constraints, sensor $\text{S}_k$, $k=1,\cdots,K$, quantizes its local observation $X_{k}$ as a discrete message $u_k\in\{0,1,\cdots,L-1\}$, with $L$ being the quantization level.
	The conditional distribution of the quantized data $u_k$ given the local observation $x_k$, i.e. $p(u_k|X_k)$, for $k=1,\cdots,K$, describes the probabilistic quantizer at sensors $k$.
	Then, the sensor transmits $u_k$ to the FC through an error free channel, and the FC receives the quantized data  $\mathbf{u}=[u_1,\cdots,u_K]^T$ from all $K$ sensors to generate the estimation of $\theta$, denoted as $\hat{\theta}(\mathbf{u})$.	
	\vspace{-10pt}

	\subsection{Problem formulation}
	To evaluate the estimation performance, we define a $\operatorname{cost}$ function 
	$C_{\theta}\left(\hat{\theta}, \mathbf{u}\right)$ for the desired parameter $\theta$, and a widely adopted one is based on the MSE principle, i.e.,
	\begin{equation}\label{eq: cost function 1}
		C_{\theta}\left(\hat{\theta}, \mathbf{u}\right)=\mathbb{E}_{\theta,\mathbf{u}}[|\theta-\hat{\theta}\left(\mathbf{u}\right)|^2].
	\end{equation}
	The optimal quantizers $\{p(u_k|X_k)\}_{k=1}^K$ and FC $\hat{\theta}(\mathbf{u})$ are chosen to minimize $C_{\theta}\left(\hat{\theta},\mathbf{u}\right)$, and the corresponding MSE minimization problem is formulated as
	\begin{equation}\label{eq: optimization problem 1}
		\underset{\left\{p(u_k|X_k)\right\}_{k=1}^{K},\,
			\hat{\theta}(\cdot)}{\min}\
		C_{\theta}\left(\hat{\theta}, \mathbf{u}\right).
	\end{equation}
	Problem \eqref{eq: optimization problem 1} is applied for various types of quantizer designs at the sensors, including the deterministic threshold quantizer and the quantizer with random dithering \cite{gray1993dithered}.
	
	\begin{comment}
	Under the considered scenario that local observations at all sensors are conditionally i.i.d., the following Lemma from \cite{6882252} shows the global optimality of utilizing identical quantizer at all sensors.
	\begin{Lemma}
	\label{Lemma: optimality of identical quantizer}
	If all sensors have conditionally i.i.d. local observations in \eqref{eq: conditionally i.i.d. local observations 1} and identical quantization level $L$, adopting identical quantizer across all sensors, i.e.,
	\begin{equation}\label{eq: identical quantizer 1}
	p(u_i=u|X_i=x)=p(u_j=u|X_j=x),
	\end{equation}
	$\forall i,j\in [K], u\in\{0,1,\cdots,L-1\}, x\in \mathbb{R}$, can  achieve the global optima for problem \eqref{eq: optimization problem 1}. 
	\end{Lemma}
\end{comment}

	For the scenario where all sensors have conditionally i.i.d. local observations and identical quantization level $L$, it was demonstrated in \cite{6882252} that adopting identical quantizer across all the sensors, i.e.,
		\begin{equation}\label{eq: identical quantizer 1}
			p(u_i=u|X_i=x)=p(u_j=u|X_j=x),
		\end{equation}
	$\forall i,j\in [K], u\in\{0,1,\cdots,L-1\}, x\in \mathbb{R}$, can  achieve the global optima for problem \eqref{eq: optimization problem 1}.
	Then, we can further prove the conditionally i.i.d. property for the quantized data from all sensors. 
	\begin{Proposition}\label{Prop: Conditional i.i.d. Proof 1}
		If all sensors have conditionally i.i.d. local observations and adopt identical quantizer, then the quantized data $\{u_k\}_{k=1}^K$ are conditionally i.i.d. with given $\theta$. In other words, we have $p(\mathbf{u}|\theta)=\prod_{k=1}^{K}p_{}(u_k|\theta)$ and 
		%\begin{equation}\label{conditional i.i.d. of U given theat 1}
			$p(u_i=u|\theta)=p(u_j=u|\theta)$,
		%\end{equation}
		$\forall i,j\in[K], u\in\{0,1\cdots,L-1\}$.
		\begin{IEEEproof}
			By subsisting \eqref{eq: conditionally i.i.d. local observations 1} and \eqref{eq: identical quantizer 1} into $p_{}(u|\theta)
			=\int_x p_{}(u|X=x)f_X(x|\theta)$ and $p(\mathbf{u}|\theta)=\prod_{k=1}^K p(u_k|\theta)$, Proposition \ref{Prop: Conditional i.i.d. Proof 1} is proved.
		\end{IEEEproof}
	\end{Proposition}

	Therefore, we ignore the indices of the sensors in the sequel, and problem \eqref{eq: optimization problem 1} can be simplified as
	\begin{equation}\label{eq: optimization problem 2}
		\underset{p(u|X),\,
			\hat{\theta}(\cdot)}{\min}\
		C_{\theta}\left(\hat{\theta}, \mathbf{u}\right).
	\end{equation}	
			
	\section{Binary Quantization Scheme}
	\label{Extensible FC neural network for binary quantizer}
	This section considers the design of identical binary probabilistic quantizer at all sensors. 
	First, the MSE lower bound for the binary-quantization-based distributed estimation is derived to serve as the benchmark for the quantization performance evaluation, and the binary probabilistic quantizer is designed to minimize this lower bound.
	Then, the optimality of the mean-fusion operation at the FC is proved, and the corresponding FC design is derived. Finally, a joint model and data driven method is proposed to sequentially train the design parameters of both the quantizer and FC modules.

	\subsection{MSE lower bound and optimality of mean-Fusion}
	\label{sec: MSE lowe bound for quantized based estimation}
	Considering the use of identical binary quantizer with $L=2$ at all sensors, i.e., $u_1,\cdots,u_K\in\{0,1\}$, the following proposition gives the achievable MSE lower bound for the binary-quantization-based distributed estimation of $\theta$ at the FC.
	\begin{Proposition}\label{MSE lower bound 1}
		When the binary quantized data $u_1,\cdots,u_K$ from all sensors are conditionally i.i.d. with given $\theta$, the MSE for estimating $\theta$ using $\theta$ is lower bounded by
		\begin{equation}
			\label{MMSE Quantizer Optimization 1}
			\begin{aligned}
				\mathbb{E}[|\theta-\hat{\theta}(\mathbf{u})|^2]
				\ge 
				\mathcal{L}_{binary}^{K}(\gamma),
			\end{aligned}
		\end{equation}
		where 
		\begin{equation}\label{eq: mse lower bound 1}
			\mathcal{L}_{binary}^{K}(\gamma)=
			\mathbb{E}[\theta^2]
			-
			\sum_{k=0}^KC_K^k
			\frac{\mathbb{E}_{\theta}^2\left[\theta \gamma(\theta)^k(1-\gamma(\theta))^{K-k}\right]}
			{\mathbb{E}_{\theta}\left[ \gamma(\theta)^k(1-\gamma(\theta))^{K-k}\right]}
		\end{equation}
		is the MSE lower bound, and
		\begin{equation}\label{eq: gamma() 1}
			\gamma(\theta)=p(u=1|\theta)=\mathbb{E}_{X}[p(u=1|X)|\theta]
		\end{equation}
		denotes the noisy quantization probability for any quantized data $u$ being "1" with given $\theta$. The equality in \eqref{MMSE Quantizer Optimization 1} holds if and only if $\hat{\theta}(\mathbf{u})=\mathbb{E}_{\theta}\left[\theta p\left(\mathbf{u}\left|\theta\right.\right)\right]/
		\mathbb{E}_{\theta}\left[ p\left(\mathbf{u}\left|\theta\right.\right)\right]$.
		\begin{IEEEproof}
			See Appendix \ref{Proof of Proposition MSE lower bound 1} for details.  	 	
		\end{IEEEproof}
	\end{Proposition}
	\begin{Remark}
	\label{Remark: MSE lower  bound 1}
	It is observed from \eqref{eq: mse lower bound 1} that $\mathcal{L}_{binary}^{K}(\gamma)$ is determined by the conditional probability distribution of the quantized data with given $\theta$, i.e.,  $p(u=1|\theta)=\gamma(\theta)$ and $p(u=0|\theta)=1-\gamma(\theta)$. With the MMSE criterion, the achievable MSE lower bound $\mathcal{L}_{binary}^{K}(\gamma)$ serves as the benchmark for the quantization performance evaluation at the sensors, and the optimal quantizer is designed to minimize $\mathcal{L}_{binary}^{K}(\gamma)$.
	\end{Remark}
	
	Proposition \ref{MSE lower bound 1} established a design metric for the quantizer by minimizing the MSE lower bound. Moreover, by demonstrating the optimality of mean-fusion operation at the FC with conditionally i.i.d. quantized data, the estimation design problem at the FC can be further simplified.
	
	\begin{Proposition}\label{Mean Value Fisher Information 1}
		If the binary quantized data $u_k$, $k=1,\cdots,K$, from all sensors are conditionally i.i.d. with given $\theta$, then the quantized data $\mathbf{u}=[u_1,\dots,u_K]^T$ and their average $\bar{u} = \frac{1}{K}\sum_{k=1}^{K} u_k$ own identical Fisher Information for any given $\theta$, i.e.,
		\begin{equation}\label{eq: fisher information 1}
			\mathbb{E}_{\mathbf{u}}\left[\frac{\partial^{2} \ln p(\mathbf{u} \mid \theta)}{\partial \theta^{2}}\right]=\mathbb{E}_{\bar{u}}\left[\frac{\partial^{2} \ln p(\bar{u} \mid \theta)}{\partial \theta^{2}}\right].
		\end{equation}	
		Furthermore, estimations of $\theta$ by using $\mathbf{u}$ and $\bar{u}$ can achieve identical MSE lower bound, i.e., 
		\begin{equation}
			\label{MMSE lower bound of mean-value U}
			\begin{aligned}
				\mathbb{E}[|\theta-\hat{\theta}(\bar{u})|^2]
				\ge
				\mathcal{L}_{binary}^{K}(\gamma).
			\end{aligned}
		\end{equation}
		where $\mathcal{L}_{binary}^{K}(\gamma)$ is given in \eqref{eq: mse lower bound 1} as the lower bound for estimation using $\mathbf{u}$. The equality in \eqref{MMSE lower bound of mean-value U} holds if and only if $\hat{\theta}(\bar{u})=\mathbb{E}_{\theta}\left[\theta p\left(\bar{u}\left|\theta\right.\right)\right]/
		\mathbb{E}_{\theta}\left[ p\left(\bar{u}\left|\theta\right.\right)\right]$.
	\end{Proposition}
	\begin{IEEEproof}
		See Appendix \ref{Proof of Proposition Mean Value Fisher Information 1} for details.	
	\end{IEEEproof}
	\begin{Remark}\label{Remark: optimality of mean fusion 1}
		According to the reciprocity of the estimation MSE and Fisher information \cite{6882252}, i.e., higher Fisher information implies lower MSE and vice versa, 
		both \eqref{eq: fisher information 1} and \eqref{MMSE lower bound of mean-value U} indicate that the achievable minimum MSE for estimating $\theta$ with either $\mathbf{u}$ or $\bar{u}$ is equivalent.
		The employment of original quantized data $\mathbf{u}$ for estimation in FC renders its input dimension susceptible to changes in the number of sensors. Therefore, when the designed FC cannot adopt to fluctuations in the number of sensors, the performance is likely to deteriorate.
		The mean-fusion operation enables the FC to use $\bar{u}$ with fixed input dimension and robustly serve the system with dynamical number of sensors in the network.
	\end{Remark}
	\vspace{-5pt}
	
\subsection{Design of probabilistic quantizer and fusion center}
\label{sec: Probabilistic quantizer design}

	According to Remarks \ref{Remark: MSE lower  bound 1} and \ref{Remark: optimality of mean fusion 1}, we are motivated to implement the binary probabilistic quantizer with random dithering and the FC with mean-fusion operation.
	\subsubsection{Probabilistic quantizer}
\begin{figure}[htbp]
	\vspace{-20pt}
	\setlength{\abovecaptionskip}{10pt}
	\setlength{\belowcaptionskip}{0pt}
	\centering
	\includegraphics[scale=0.85]{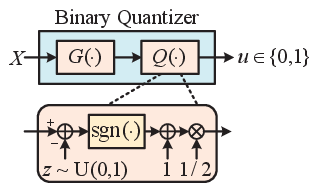}
	\caption{Binary probabilistic quantizer}
	\label{Uniform Dithering realization of probabilistic quantization}
\end{figure}

	As shown in Fig. \ref{Uniform Dithering realization of probabilistic quantization}, we consider one implementation of the binary probabilistic quantizer design with random dithering. Define the probability controller $G(\cdot):\mathbb{R}\rightarrow [0,1]$. The local observation $X$ is first sent to $G(\cdot)$, and the output $G(X)$ is then fed into the quantization function $Q(\cdot):[0,1]\rightarrow\{0,1\}$ to generate a random binary data
	\begin{equation}\label{eq: probabilistic quantization 1}
		u=Q(G(X))=\frac{1+\mathrm{sgn}(G(X)-z)}{2}\in\{0,1\},
	\end{equation}
	where $z\sim\mathrm{U}(0,1)$ is a standard uniform distributed dithering noise, and $\mathrm{sgn}(\cdot)$ is the sign function.
	
	It is observed from $\eqref{eq: probabilistic quantization 1}$ that the probability of the local observation $X$ being quantized as $u=1$ is equal to $G(X)$, i.e.,
	\begin{equation}\label{eq: uniform dithering realization 1}
	\begin{aligned}
		p(u=1|X)
		=&p\left(\frac{1+\mathrm{sgn}(G(X)-z)}{2}=1\Big|X\right)\\
		=&p(G(X)>z|X)\\
		=&G(X), 
	\end{aligned}
	\end{equation}
	and intuitively we have $p(u=0|X)=1-G(X)$.	
	Therefore, by using Proposition \ref{MSE lower bound 1} and \eqref{eq: uniform dithering realization 1}, to minimize the MSE lower bound $\mathcal{L}_{binary}^{K}(\gamma)$ in \eqref{eq: mse lower bound 1} is equivalent to find the optimal probability controller $G(\cdot)$ for the quantizer, i.e.,
	\begin{equation}
		\label{noise case: minimize mmse 3}
		\underset{G(\cdot)}{\min}\quad 
		\mathcal{L}_{binary}^{K}(\gamma),
	\end{equation}%
where $\gamma(\theta)$ defined in \eqref{eq: gamma() 1} is rewritten as
\begin{equation}\label{eq: gamma() 2}
	\begin{split}
		\gamma(\theta)
		= \mathbb{E}_{X}[G(X)|\theta].
	\end{split}
\end{equation}

%%%%%%%%%%%%%%%%%%%%%%%%%%%%%%%%
\begin{figure*}[!b]
	\normalsize
	\newcounter{mytempeqncnt}
	\setcounter{mytempeqncnt}{\value{equation}}
	\setcounter{equation}{18}
	\hrulefill
	\begin{align}
		\label{loss function: 1}
		\underset{\Phi}{\min}\ \ &
		\mathcal{L}_{binary}^K(\Phi)=
		\mathbb{E}[\theta^2]-\sum_{k=0}^KC_K^k
		\left({\mathbb{E}_{\theta}^2\left[\theta \left(\gamma_{\Phi}(\theta)\right)^k
			\left(1-\gamma_{\Phi}(\theta)\right)^{K-k}\right]}\right)/
		\left({\mathbb{E}_{\theta}\left[ \left(\gamma_{\Phi}(\theta)\right)^k
			\left(1-\gamma_{\Phi}(\theta)\right)^{K-k}\right]}\right),\\
		\label{loss function: 2}
		\underset{\{\Phi,\Psi\}}{\min}\ &
		\mathcal{T}_{binary}^K(\Phi,\Psi)=
		\sum_{k=0}^{K}C_{K}^{k}\mathbb{E}_{\theta}\left[\left|\theta-F_{\Psi}\left(\bar{u}=\frac{k}{K}\right)\right|^2(\gamma_{\Phi}(\theta))^k\left(1-\gamma_{\Phi}(\theta)\right)^{K-k}\right].
	\end{align}
	\setcounter{equation}{\value{mytempeqncnt}}
	%\hrulefill
	\vspace*{4pt}
\end{figure*}
\subsubsection{Fusion center}
\begin{figure}[htbp]
	\vspace{-10pt}
	\setlength{\abovecaptionskip}{5pt}
	\setlength{\belowcaptionskip}{-5pt}
	\centering
	\includegraphics[scale=0.85]{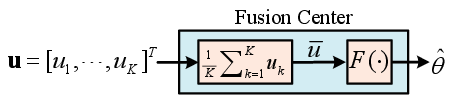}
	\caption{Mean-fusion operation for binary-quantization-based distributed estimation.}
	\label{Mean value fusion for one-bit quantizer}
\end{figure}

As shown in Fig \ref{Mean value fusion for one-bit quantizer}, we are motivated by Remark \ref{Remark: optimality of mean fusion 1} to implement a FC with mean-fusion operation. After the binary quantized data $\{u_k\}_{k=1}^K$ from all sensors are received by the FC, they are first averaged to get $\bar{u} = \frac{1}{K}\sum_{k=1}^{K}u_k$.
Then, the desired parameter $\theta$ is estimated as 
\begin{equation}\label{eq: estimator function 1}
	\hat{\theta}=F(\bar{u}),
\end{equation} 
where $F(\cdot){:[0,1]\rightarrow \mathbb{R}}$ is the estimator function to be designed.
From \eqref{eq: estimator function 1} and by using 
$$p\left(\bar{u}=\frac{k}{K}\big|\theta\right)=C_K^k\left(\gamma(\theta)\right)^k\left(1-\gamma(\theta)\right)^{K-k},$$ 
which is derived in \eqref{eq: bernoulli binomial distribution 2}, the estimation MSE for $\theta$ at the FC is computed as
\begin{equation}
	\label{minimize mmse 3}
	\begin{aligned}
		\mathcal{T}_{binary}^{K}(F)
		=&\
		\mathbb{E}_{\theta,\bar{u}}[|\theta-F(\bar{u})|^2]\\
		%=&
		%\mathbb{E}_{\theta}\left[\mathbb{E}_{\bar{u}}\left[\left|\theta-%F(\bar{u})\right|^2\big|\theta\right]\right]\\
		=& \mathbb{E}_{\theta}\left[\sum_{k=0}^{K}\left|\theta-F\left(\frac{k}{K}\right)\right|^2p\left(\bar{u}=\frac{k}{K}\bigg|\theta\right)\right]\\
		=& \sum_{k=0}^{K}C_{K}^{k}\mathbb{E}_{\theta}\left[\left|\theta-F\left(\frac{k}{K}\right)\right|^2\gamma(\theta)^k\left(1-\gamma(\theta)\right)^{K-k}\right].
	\end{aligned}
\end{equation} 
For the goal to minimize the estimation MSE in \eqref{minimize mmse 3}, the best estimator function for FC is designed as
\begin{equation}
	\label{minimize mmse 4}
	\underset{F(\cdot)}{\min}\
	\mathcal{T}_{binary}^K(F).
\end{equation}%

\begin{Remark}\label{Remark: optimization of G() 1}
	If perfect statistic knowledge on the distributions of the desired parameter and noise is available, problems \eqref{noise case: minimize mmse 3} and \eqref{minimize mmse 4} are variational problems of $G(\cdot)$ and $F(\cdot)$, which can be solved by calculus of variations under certain conditions \cite{6882252}. The non-closed-form solution of $G(\cdot)$ can also be obtained based on the parametric and data driven method with data samples generated from the respective distributions\cite{6174480}.
	%However, the closed-form solution of optimal $G(\cdot)$ for problem \eqref{noise case: minimize mmse 3} is difficult to compute due to the non-closed-form expectation and integral in \eqref{noise case: minimize mmse 3} and \eqref{eq: gamma() 2},
	Since perfect information about the distributions of the desired parameter and observation noise is difficult to be obtained in practical systems, it is more interesting to study the scenarios that both the above distributions are unknown or only the noise distribution is known. Thus, a joint model and data driven method is proposed as shown in the next subsection.
\end{Remark}

\subsection{Joint model and data driven method}
\label{sec: Neural Probabilistic Quantizer Training}
A joint model and data driven method to solve problems \eqref{noise case: minimize mmse 3} and \eqref{minimize mmse 4} is proposed in the following:
\subsubsection{Binary quantizer}
\begin{figure}[htbp]
	\centering
	\vspace{0pt}
	\setlength{\abovecaptionskip}{0pt}
	\setlength{\belowcaptionskip}{-10pt}
	%\subfigtopskip=0pt 
	%\subfigbottomskip=15pt 
	%\subfigcapskip=5pt 
	\subfloat[Probability controller DNN]{
		%\label{noise_case_multi_bit 1: subfig1}
		\includegraphics[scale=0.8]{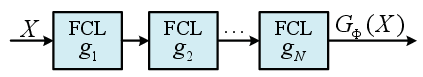}}
	\\
	\subfloat[Estimator DNN]{
		%\label{noise_case_multi_bit 1: subfig2}
		\includegraphics[scale=0.8]{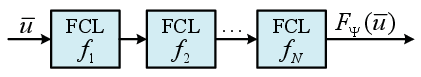}}
	\vspace{10pt}
	\caption{Architecture of the probability controller and estimator DNNs.}
	\label{Fig: architecture of the DNNs}
\end{figure}

For the binary probabilistic quantizer defined in  Fig. \ref{Uniform Dithering realization of probabilistic quantization}, its probability controller $G(\cdot)$ is implemented as a DNN $G_{\Phi}$ with $N$ fully connected layers (FCLs)\cite{o2017introduction}.
In the DNN $G_{\Phi}$ as shown in Fig. \ref{Fig: architecture of the DNNs}(a), the input observation $X$ is fed into $N$ FCLs, i.e.,
\begin{equation}
	G_{\Phi}(X)=g_N\circ\cdots\circ g_2\circ g_1(X),
\end{equation}
where $g_i$, $i\in[N]$, is the FCL with the parameter $\alpha_N$, the number of input dimensions being $L_{g,I,i}$, and the number of output dimensions being $L_{g,O,i}$. Note that $L_{g,I,1}=L_{g,O,N}=1$ is fixed. 
To mitigate the issues of gradient explosion and vanishing, the ReLU function \cite{o2017introduction} is employed as the activation function for $g_1, g_2, \cdots, g_{N-1}$. In the output layer $g_N$, the Sigmoid function is utilized as the activation function to ensure the output range is confined to $[0,1]$, as it represents the quantization probability.
To summarize, the training parameter of $G_{\Phi}$ is $\Phi=\{\alpha_1,\alpha_2,\cdots,\alpha_N\}$.
\subsubsection{FC}

For the FC defined in Fig. \ref{Mean value fusion for one-bit quantizer}, its estimator function $F(\cdot)$ is implemented as a DNN $F_{\Psi}$ with $N$ FCLs, denoted as $f_1, f_2, \cdots, f_N$. Similarly, the ReLU function is employed as the activation function for $f_1, f_2, \cdots, f_{N-1}$ to address gradient-related challenges. In the output layer $f_N$, the Tanh function is utilized as the activation function, as it ensures the output of $F_{\Psi}$ represents a normalized estimator within the range of $[-1,1]$.
The training parameter of  $F_{\Psi}$ is denoted as $\Psi=\{\beta_1,\beta_2,\cdots,\beta_N\}$, where $\beta_i$ is the parameter of the FCL $f_i$, $i\in[N]$. $L_{f,I,i}$ and $L_{f,O,i}$ are the numbers of input dimensions and output dimensions of the FCL $f_i$, $i\in[N]$, with $L_{f,I,1}=L_{f,O,N}=1$ being fixed.

Then, by adopting the DNNs, optimization problems \eqref{noise case: minimize mmse 3} and \eqref{minimize mmse 4} are rewritten as \eqref{loss function: 1} and \eqref{loss function: 2}.
\begin{Remark}
Based on the MMSE criterion, problems \eqref{loss function: 1} and \eqref{loss function: 2} imply that the optimal quantizer design is independent of the FC design and the optimal FC design is obtained based on the optimal quantizer. For the practical scenario that the quantizer and FC are separated in space, the regular joint training method of the two modules
a sequential deep learning training method to obtain optimal $G_{\Phi}$ and $F_{\Psi}$ is proposed in the next subsection. Notice that the sequential training 	
\end{Remark}

\vspace{-5pt}
\subsection{Sequential training method}
\label{sec: Sequential training of binary quantizer and FC}
For the sequential training of the probability controller DNN $G_{\Phi}$ in quantizer and estimator DNN $F_{\Psi}$ in FC, we aim to first find the optimal parameter $\Phi^{*}$ which minimizes the loss function $\mathcal{L}_{binary}^K(\Phi)$ in \eqref{loss function: 1}, and then find the optimal $\Psi^{*}$ that minimizes the loss function $\mathcal{T}_{binary}^K(\Phi^*,\Psi)$ in \eqref{loss function: 2}, 
under two cases that the distributions of both the desired parameter and observation noise are unknown or only the noise distribution is known. 
Besides, it is observed that the two loss functions  are contingent upon the number of sensors $K$. Given the consideration of a practical network where the number of sensors may vary dynamically, a predetermined number of sensors is chosen during the training phase, while the efficacy of the trained model is assessed using varying sensor quantities during the testing phase.

\subsubsection{Training with unknown parameter and noise distributions}
The training process is based on the data set $D_1=\{\theta_t,\mathbf{x}_t\}_{t=1}^T$, where $\theta_t$ is the $t$-th sample of the desired parameter $\theta$ to be estimated, $\mathbf{x}_t=\{x_{t,1},\cdots,x_{t,Q}\}$ contains $Q$ noise-corrupted observation samples of $\theta_t$, 
and $T$ denotes the total number of training samples. The desired parameter $\theta_t$ is obtained from the experimental environment, and the observation 
$x_{t,1},\cdots,x_{t,Q}$
are obtained from the sensor by periodically observing $\theta_t$ under the same environment.

At each epoch, based on the mini batch method\cite{o2017introduction}, the whole data set is divided into $T/B$ batches
where $B$ is the number of batch samples. Here, we consider the case that $T/B$ is an integer, without loss of generality. In the exceptional scenario, a simple approach is to randomly select additional samples from the existing dataset and append them to form a new dataset that satisfies the requirement. Parameter $\Phi$ is trained for $T/B$ times within an epoch, where each time a new batch set is utilized for training. 
Within each time, the loss function in \eqref{loss function: 1} is approximated and averaged on the whole batch samples as 
\setcounter{equation}{20}
\begin{equation}
	\label{loss fucntion: 11}
	%\begin{aligned}
	\hat{\mathcal{L}}_1=
	\sum_{t=1}^B(\theta_t)^2
	-\sum_{k=0}^{K_S}C_{K_S}^{k}
	\frac{\left(\sum_{t=1}^B
		\theta_t\left(
		\gamma_{\Phi}^t\right)^{K_S}
		\left(1-\gamma_{\Phi}^t\right)^{{K_S}-k}\right)^2}
	{\sum_{t=1}^B\left(
		\gamma_{\Phi}^t\right)^{K_S}
		\left(1-\gamma_{\Phi}^t\right)^{{K_S}-k}},
	%\end{aligned}
\end{equation}
where $K_S$ is the predetermined sensor quantity parameter in the training and
\begin{equation}\label{approximation of Phi(theta) 1}
	\gamma_{\Phi}^t=\sum_{q=1}^QG_{\Phi}(x_{t,q})
\end{equation}
is the empirical approximation of $\gamma_{\Phi}(\theta_t)=\mathbb{E}_{X}[G_{\Phi}(X)|\theta_t]$ over data set $\{\theta_t,\mathbf{x}_t\}$. 
By using the back propagation algorithm\cite{o2017introduction}, the gradient $\nabla_{\Phi}\hat{\mathcal{L}_1}$ is calculated based on \eqref{loss fucntion: 11}, and the parameter $\Phi$ is updated epoch by epoch. After the maximum  number of training epochs is reached, the optimal parameter is obtained as $\Phi^*$. 

Once the optimal probability controller $G_{\Phi^*}$ in the quantizer is obtained, it is utilized to sequentially train the estimator DNN $F_{\Psi}$ in the FC. The training of $F_{\Psi}$ uses the same mini batch method based on data set $D_1$. 
Within each time, the loss function in \eqref{loss function: 2} is approximated and averaged on the whole batch samples as
\begin{equation}
	\label{loss fucntion: 21}
	\begin{aligned}
		\hat{\mathcal{T}}_1
		=&\sum_{t=1}^{B}\sum_{k=0}^{K_F}
		\left(\theta_t-F_{\Psi}\left(\frac{k}{K_F}\right)\right)^2
		C_{K_F}^{k}\left(\gamma_t^*\right)^k
		\left(1-\gamma_t^*\right)^{{K_F}-k},
	\end{aligned}
\end{equation}
where $K_F$ is the predetermined sensor quantity parameter in the training and
\begin{equation}\label{approximation approximation of Phi(theta) 2}
	\gamma_t^*= \sum_{q=1}^QG_{\Phi^*}(x_{t,q}).
\end{equation}
Based on \eqref{loss fucntion: 21}, parameter $\Psi$ is updated epoch by epoch using the back propagation algorithm\cite{o2017introduction}, and optimal $\Psi^*$ is obtained once the maximum training epochs are reached.

\subsubsection{Training with known noise distribution}
If the statistic information about the distribution of the observation noise is available, then intuitively  $f_X({\cdot|\theta})$ is obtained and we can use only data set $D_2=\{\theta_t\}_{t=1}^T$ for the sequential training.
Considering the case that the sensor's observation range is restricted to $[-W,W]$, we define an artificial observation set $O=\{-W,-\frac{Q-1}{Q}W,\cdots,0,\frac{1}{Q}W,\cdots,W\}$
to cover the bounded observations.
Based on the information of distribution $f_X(\cdot|\theta)$, the loss function $\mathcal{L}_{binary}^K(\Phi)$ in \eqref{loss function: 1} is approximated based on the whole batch samples as
\begin{equation}
	\label{loss fucntion: 12}
	%\begin{aligned}
	\hat{\mathcal{L}}_2=
	\sum_{t=1}^B(\theta_t)^2
	-\sum_{k=0}^{K_S}C_{K_S}^k
	\frac{\left(\sum_{t=1}^B
		\theta_t\left(
		\gamma_{\Phi}^t\right)^k
		\left(1-\gamma_{\Phi}^t\right)^{{K_S}-k}\right)^2}
	{\sum_{t=1}^B\left(\gamma_{\Phi}^t\right)^k
		\left(1-\gamma_{\Phi}^t\right)^{{K_S}-k}},
	%\end{aligned}
\end{equation} 
where $\gamma_{\Phi}^t=\sum_{x\in O}G_{\Phi}(x)f_X(x|\theta_t)$.
By using the back propagation algorithm, parameter $\Phi$ is updated epoch by epoch based on \eqref{loss fucntion: 12}, and optimal $\Phi^*$ is obtained after the maximum  number of training epochs is reached.

With the optimal probability controller $G_{\Phi^*}$, the estimator DNN $F_{\Psi}$ in the FC is sequentially trained. The training of FC utilizes the same mini batch method based on the sets $D_2$ and $O$. 
Within each time, the loss function in \eqref{loss function: 2} is approximated and averaged on the whole batch samples as
\begin{equation}
	\label{loss fucntion: 22}
	\begin{aligned}
		\hat{\mathcal{T}}_2
		=&\sum_{t=1}^{B}\sum_{k=0}^{K_F}
		\left(\theta_t-F_{\Psi}\left(\frac{k}{K_F}\right)\right)^2
		C_{K_F}^k\left(\gamma_t^*\right)^{K_F}
		\left(1-\gamma_t^*\right)^{{K_F}-k},
	\end{aligned}
\end{equation}
where $\gamma_t^*= \sum_{x\in O}G_{\Phi^*}(x)f_X(x|\theta_t)$. 
Based on \eqref{loss fucntion: 22}, optimal parameter $\Psi^*$ is obtained by updating $\Psi$ epoch by epoch using the back propagation algorithm until the maximum number of training epochs are reached.

\begin{figure}[htbp]
	\vspace{0pt}
	\setlength{\belowcaptionskip}{-10pt}
	\centering
	\includegraphics[width=230pt]{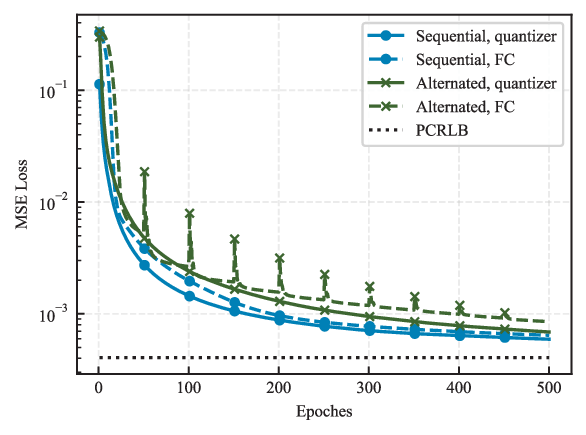}
	\caption{Convergence of the quantizer and FC training.}
	\label{Fig: Comparsion_of_loss 1}
\end{figure}

Finally, the performance of the proposed sequential training scheme is compared with an canonical alternative training scheme, in which a total of 500 epochs are divided into 10 rounds, and the quantizer and FC DNNs are trained alternately for each round. Figure \ref{Fig: Comparsion_of_loss 1} illustrates the plotted MSE loss of both the quantizer and FC with respect to the training epochs. Notably, the proposed sequential training scheme demonstrates superior and smoother convergence during the training phase. Additionally, the final MSE loss achieved by the sequential training scheme is smaller compared to that of the alternated training approach. This strongly supports the effectiveness of the proposed sequential training method.

%%%%%%%%%%%%%%%%%%%%%%%%%%%%%%%%%%%%%%%%%%%%%%%%%%%
\begin{figure*}[!b]
	\normalsize
	\newcounter{mytempeqncnt2}
	\setcounter{mytempeqncnt2}{\value{equation}}
	\setcounter{equation}{32}
	\hrulefill
	\begin{equation}\label{eq: parallel MSE cost 1}
		\begin{aligned}
			\mathcal{T}_{\mathrm{parallel}}^{M,K}(\Phi,\Psi)
			=
			\mathbb{E}_{\theta}\left[\sum_{\bar{\mathbf{u}}}\left|\theta-F_{\Psi}\left(\bar{\mathbf{u}}\right)\right|^2p\left(\bar{\mathbf{u}}|\theta\right)\right]
			= 
			\sum_{i_1=0}^{K}\cdots\sum_{i_{M}=0}^{K}
			\mathbb{E}_{\theta}\left[\left|\theta-F_{\Psi}\left(\frac{i_1,\cdots,i_M}{K}\right)\right|^2p_{\Phi}^{i_1,\cdots,i_M}(\theta)\right].
		\end{aligned}
	\end{equation}
	\setcounter{equation}{\value{mytempeqncnt2}}
	%\hrulefill
	\vspace*{4pt}
\end{figure*}
\section{Multi-bit Probabilistic Quantization}
\label{sec: Multi-bit Probabilistic Quantizer and Fusion Center Design for MMSE Estimation}
In this section, we relax the assumption of one-bit quantization constraint at sensors and address the optimal design of multi-bit probabilistic quantizer and FC. 
We consider two different joint model and data driven multi-bit quantization schemes, corresponding to parallel and one-hot implementation of quantization for all bits in a multi-bit quantized information. 
Owing to the similarities in the optimization of the multi-bit quantizer and FC design with that under the binary quantization constraint discussed in Section III, we omit the details of the deep learning training process for the quantizer and FC in this section.

\subsection{Parallel quantization}
\begin{figure*}[htp]
	\normalsize
	\centering
	\vspace{0pt}
	\setlength{\abovecaptionskip}{5pt}
	\setlength{\belowcaptionskip}{0pt}
	%\subfigtopskip=0pt 
	%\subfigbottomskip=15pt 
	%\subfigcapskip=5pt 
	\subfloat[Quantizer]{
		%\label{noise_case_multi_bit 1: subfig1}
		\includegraphics[scale=0.8]{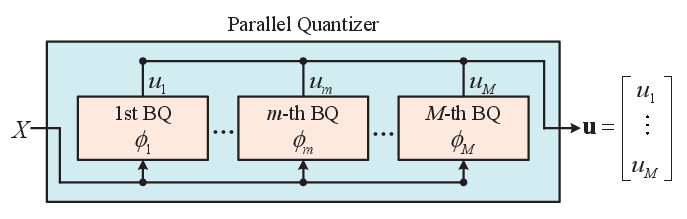}}\quad\quad\quad\quad
	\subfloat[FC]{
		%\label{noise_case_multi_bit 1: subfig2}
		\includegraphics[scale=0.8]{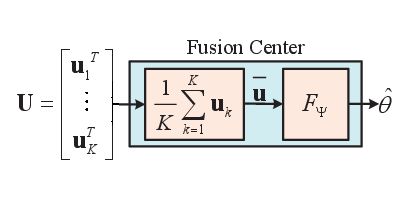}}
	\vspace{4pt}
	\caption{Multi-bit parallel quantization and mean-vector-fusion.}
	\label{noise_case_multi_bit 1}
\end{figure*}
The joint model and data driven multi-bit parallel quantizer and FC modules are shown in Fig. \ref{noise_case_multi_bit 1}:
\begin{itemize}
\item Quantizer: As shown in Fig. \ref{noise_case_multi_bit 1}(a), we consider an $M$-bit parallel quantizer module where the quantization is implemented by $M$ parallel binary quantizers (BQ). Each BQ adopts identical binary quantizer design shown in Fig. \ref{Uniform Dithering realization of probabilistic quantization}. Taking the $m$-th BQ as an example, it maps the input observation $X$ into a binary output $u_m\in\{0,1\}$ following the distribution $p(u_m=1|X)=G_{\phi_m}(X)$ and $p(u_m=0|X)=1-G_{\phi_m}(X)\}$, where $G_{\phi_m}$ is the probability controller DNN in the $m$-th BQ with design parameter $\phi_m$. Thus, $M$ BQs map the observation as an $M$-bit quantization message $\mathbf{u}=[u_1,u_2,\cdots,u_M]^T\in\{0,1\}^M$.

\item FC: As shown in Fig. \ref{noise_case_multi_bit 1}(b), we consider the scenario that $K$ sensors adopt identical $M$-bit parallel quantizer with design parameters $\Phi=\{\phi_1,\cdots,\phi_m\}$. The FC receives $K$ quantized messages $\mathbf{U}=[\mathbf{u}_1,\cdots,\mathbf{u}_K]^T$ from $K$ sensors and uses the mean-vector-fusion operation to obtain $\bar{\mathbf{u}}=\frac{1}{K}\sum_{k=1}^K\mathbf{u}_k$. The desired parameter $\theta$ is estimated as
\begin{equation}\label{eq: parallel estimator function 1}
	\hat{\theta}=F_{\Psi}(\bar{\mathbf{u}}),
\end{equation}
where  $F_{\Psi}(\cdot):[0,1]^M\rightarrow \mathbb{R}$ 
is the estimator DNN with design parameters $\Psi$.
\end{itemize}

Similar to Proposition \ref{MSE lower bound 1}, the following proposition shows that if identical multi-bit parallel quantizer is deployed at all sensors, then adopting the mean-vector-fusion operation on the quantized data causes no performance degradation for the estimation of the desired parameter.

\begin{Proposition}\label{Prop: Paraller quantization 1}
If all sensors adopt the identical $M$-bit parallel quantizer with design parameter $\Phi=\{\phi_1,\cdots,\phi_m\}$, estimation of $\theta$ by using the original quantized data matrix $\mathbf{U}$ or only the mean-vector $\bar{\mathbf{u}}$ can achieve identical MSE lower bound, i.e.,
\begin{align}
	\label{eq: minimize mmse 7 U}
		\mathbb{E}[|\theta-\hat{\theta}(\mathbf{U})|^2]
		\ge&
		\mathcal{L}_{\mathrm{parallel}}^{M,K}(\Phi),\\
	\label{eq: minimize mmse 7 u}
		\mathbb{E}[|\theta-\hat{\theta}(\bar{\mathbf{u}})|^2]
		\ge&
		\mathcal{L}_{\mathrm{parallel}}^{M,K}(\Phi),
\end{align}
where 
\begin{equation}\label{eq: parallel MSE lower bound 1}
	\mathcal{L}_{\mathrm{parallel}}^{M,K}(\Phi)=\sum_{i_1=0}^{K}
	\cdots
	\sum_{i_{M}=0}^{K}
	\frac{\mathbb{E}_{\theta}^2\left[\theta  p_{\Phi}^{i_1,\cdots,i_M}(\theta)\right]}
	{\mathbb{E}_\theta\left[p_{\Phi}^{i_1,\cdots,i_M}(\theta)\right]},
\end{equation}
is the MSE lower bound,
\begin{equation}
	\begin{aligned}
	p_{\Phi}^{i_1,\cdots,i_M}(\theta)
	=&p\left(\bar{\mathbf{u}}=\frac{1}{K}[i_1,\cdots,i_M]^T\Big|\theta\right)\\
	=&\prod_{m=1}^MC_{K}^{i_m}(\gamma_{\phi_m}(\theta))^{i_m}(1-\gamma_{\phi_m}(\theta))^{K-i_m},
	\end{aligned}
\end{equation}
and
\begin{equation}\label{eq: parallel mean distribution 1}
\gamma_{\phi_m}(\theta)=p([\mathbf{u}]_m=1|\theta)
=\mathbb{E}_X[G_{\phi_m}(X)|\theta]
\end{equation}
denotes the conditional probability for the $m$-th bit of any quantized vector being "$1$" with given $\theta$. The equality in \eqref{eq: minimize mmse 7 U} holds if and only if $\hat{\theta}(\mathbf{U})=\mathbb{E}_{\theta}\left[\theta p\left(\mathbf{U}\left|\theta\right.\right)\right]/
\mathbb{E}_{\theta}\left[ p\left(\mathbf{U}\left|\theta\right.\right)\right]$, and the equality in \eqref{eq: minimize mmse 7 u} holds if and only if $\hat{\theta}(\bar{\mathbf{u}})=\mathbb{E}_{\theta}\left[\theta p\left(\bar{\mathbf{u}}\left|\theta\right.\right)\right]/
\mathbb{E}_{\theta}\left[ p\left(\bar{\mathbf{u}}\left|\theta\right.\right)\right]$.

\begin{IEEEproof}
	See Appendix \ref{Proof of Proposiiton {Prop: Paraller quantization 1}} for details.
\end{IEEEproof}
\end{Proposition}

By using \eqref{eq: parallel estimator function 1} and \eqref{eq: parallel mean distribution 1} and following the analysis in \eqref{minimize mmse 3}, the estimation MSE for $\theta$ at the FC is computed as \eqref{eq: parallel MSE cost 1}.
From \eqref{eq: parallel MSE lower bound 1} and \eqref{eq: parallel MSE cost 1}, the optimization problems for the multi-bit parallel quantizer and FC are derived as
\setcounter{equation}{33}
\begin{align}
	&\underset{\Phi}{\min}\ \mathcal{L}_{\mathrm{parallel}}^{M,K}(\Phi),\\
	&\underset{\{\Phi,\Psi\}}{\min}\ \mathcal{T}_{\mathrm{parallel}}^{M,K}(\Phi,\Psi).
\end{align}
Since the deep learning based training for $\Phi$ and $\Psi$ is similar to that for the binary quantization scenario discussed in section \ref{sec: Sequential training of binary quantizer and FC}, we ignore the details of the training process.

\begin{figure*}[!b]
	\normalsize
	\newcounter{mytempeqncnt31}
	\setcounter{mytempeqncnt31}{\value{equation}}
	\hrulefill
	%%%
	\setcounter{equation}{38}
	\begin{equation}\label{eq: joint MSE lower bound 1}
		\mathcal{L}_{\mathrm{onehot}}^{M,K}(\Phi)=
		\mathbb{E}[\theta^2]-
		\sum_{i_0=0}^{K}\cdots\sum_{i_k=0}^{K-\sum\limits_{l=0}^{k-1}i_l}\cdots\sum_{i_{L-2}=0}^{K-\sum\limits_{l=0}^{L-3}i_l}
		\left({\mathbb{E}_{\theta}^2\left[\theta q_{\Phi}^{i_0,\cdots,i_{L-1}}(\theta)\right]}\right)/
		\left({\mathbb{E}_{\theta}\left[q_{\Phi}^{i_0,\cdots,i_{L-1}}(\theta)\right]}\right),
	\end{equation}
	%%%%
	\setcounter{equation}{40}
	\begin{equation}\label{eq: joint MSE cost 1}
		\begin{aligned}
			\mathcal{T}_{\mathrm{onehot}}^{M,K}(\Phi,\Psi)
			= \mathbb{E}_{\theta}\left[\sum_{\bar{\mathbf{v}}}\left|\theta-F_{\Psi}\left(\bar{\mathbf{v}}\right)\right|^2p\left(\bar{\mathbf{v}}|\theta\right)\right]
			=
			\sum_{i_0=0}^{K}\cdots\sum_{i_k=0}^{K-\sum\limits_{l=0}^{k-1}i_l}\cdots\sum_{i_{L-2}=0}^{K-\sum\limits_{l=0}^{L-3}i_l}
			\mathbb{E}_{\theta}\left[\left|\theta-F_{\Psi}\left(\frac{i_0,\cdots,i_{L-1}}{K}\right)\right|^2q_{\Phi}^{i_0,\cdots,i_{L-1}}(\theta)\right].
		\end{aligned}
	\end{equation}
	%%%%
	\setcounter{equation}{\value{mytempeqncnt31}}
	%\hrulefill
	\vspace*{4pt}
\end{figure*}
\subsection{One-hot quantization}
\begin{figure*}[htp]
	\normalsize
	\centering
	\vspace{0pt}
	\setlength{\abovecaptionskip}{0pt}
	\setlength{\belowcaptionskip}{-10pt}
	%\subfigtopskip=0pt 
	%\subfigbottomskip=15pt 
	%\subfigcapskip=5pt 
	\subfloat[Quantizer]{
		%\label{noise_case_multi_bit 1: subfig1}
		\includegraphics[scale=0.8]{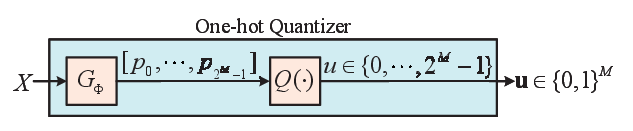}}\quad\quad\quad\quad
	\subfloat[FC]{
		%\label{noise_case_multi_bit 1: subfig2}
		\includegraphics[scale=0.8]{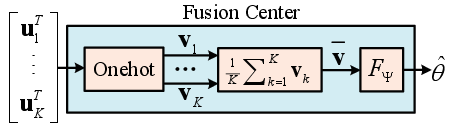}}
	\vspace{10pt}
	\caption{Multi-bit one-hot quantization and mean-one-hot-vector fusion.}
	\label{noise_case_multi_bit 2}
\end{figure*}
The joint model and data driven multi-bit one-hot quantizer and FC modules are shown in Fig. \ref{noise_case_multi_bit 2}:
\begin{itemize}
\item
Quantizer: 
As shown in Fig. \ref{noise_case_multi_bit 2}(a), we consider an $M$-bit one-hot quantization module where the probability distribution of the $M$-bit quantized data is controlled by an one-hot probability controller DNN $G_{\Phi}(\cdot):\mathbb{R}\rightarrow [0,1]^{2^M}$ with softmax output and design parameter $\Phi$.
Local observation $X$ is first sent to $G_{\Phi}$ and mapped as a $2^M$-dimensional probability vector $\mathbf{p}=[p_0,\cdots,p_{2^M-1}]^T=G_{\Phi}(X)$, where $0\leq p_m\leq 1$ and $\sum_{m=0}^{2^M-1}p_m=1$.
Then, the probability vector $\mathbf{p}$ is fed into the  quantization function $Q(\cdot):[0,1]^{2^M}\rightarrow\{0,1,\cdots,2^M-1\}$ to generate a quantized value $u=Q(\mathbf{p})$. Similar to the quantization function $Q(\cdot)$ in binary quantizer, function $Q(\cdot)$ in $M$-bit one-hot quantizer is to make the probability distribution $\{p(u=m|X)\}_{m=0}^{2^M-1}$ being controlled by the probability vector $\mathbf{p}$, i.e.,
\begin{equation}
	p(u=m|X)=p(Q(\mathbf{p})=m|X)=p_m=[G_{\Phi}(X)]_m,
\end{equation}
for $m=0,1,\cdots,2^M-1$, where $[G_{\Phi}(X)]_m$ is the $m$-th entry of vector $G_{\Phi}(X)$. Then, the quantized value $u$ is transformed into a corresponding binary vector  $\mathbf{u}\in\{0,1\}^M$, which is then being transmitted to the FC.

\item
FC: As shown in Fig. \ref{noise_case_multi_bit 2}(b), we consider the scenario that $K$ sensors adopt the identical $M$-bit one-hot quantizer with design parameter $\Phi$. 
The FC receives $K$ quantized messages $\mathbf{U}=[\mathbf{u}_1,\cdots,\mathbf{u}_K]^T$ from $K$ sensors  
and transforms them into $K$ one-hot vectors as $\mathbf{V}=[\mathbf{v}_1, \cdots, \mathbf{v}_K]^T$. 
Based on the one-hot encoding scheme \cite{o2017introduction}, each vector $\mathbf{u}_k\in\{0,1\}^M$ is transformed as a $2^M$-dimensional vector $$\mathbf{v}_k=
\begin{cases}
	[1,\mathbf{0}_{2^M-1}^T]^T, &(\mathbf{u}_k)_{10}=0,\\
	[\mathbf{0}_{(\mathbf{u}_k)_{10}}^T,1,\mathbf{0}_{2^M-(\mathbf{u}_k)_{10}-1}^T]^T, &(\mathbf{u}_k)_{10}\ge 1,\\
\end{cases}$$
where $\mathbf{0}_{2^M-1}^T$ is $(2^M-1)$-dimensional zero vector and $(\mathbf{u}_k)_{10}$ is the decimal expression of $\mathbf{u}_k$.
Then, the mean-one-hot-vector $\bar{\mathbf{v}}=\frac{1}{K}\sum_{k=1}^K\mathbf{v}_k$ is utilized to estimate the desired parameter $\theta$ as $\hat{\theta}=F_{\Psi}(\bar{\mathbf{v}})$, 
where $F_{\Psi}(\cdot):[0,1]^{2^M}\rightarrow \mathbb{R}$
is the estimator DNN with design parameter $\Psi$.
\end{itemize}

The following proposition shows that if identical multi-bit one-hot quantizer is deployed at all sensors, then adopting the mean-one-hot-vector-fusion operation on the quantized data causes no performance degradation for the estimation of the desired parameter.

\begin{Proposition}\label{Prop: Joint quantization 1}
If all sensors adopt the identical $M$-bit one-hot quantizer with design parameter $\Phi$, estimation of $\theta$ by using the original quantized data matrix $\mathbf{U}$ or only the mean-one-hot-vector $\bar{\mathbf{v}}$ can achieve identical MSE lower bound, i.e.,	
	\begin{equation}\label{minimize mmse 8}
	\begin{aligned}
		\mathbb{E}[|\theta-\hat{\theta}(\mathbf{U})|^2]
		\ge
		\mathcal{L}_{\mathrm{onehot}}^{M,K}(\Phi),
\end{aligned}
\end{equation}
\begin{equation}\label{minimize mmse 9}
	\begin{aligned}
		\mathbb{E}[|\theta-\hat{\theta}(\bar{\mathbf{v}})|^2]
		\ge
		\mathcal{L}_{\mathrm{onehot}}^{M,K}(\Phi),
	\end{aligned}
\end{equation}
where $\mathcal{L}_{\mathrm{onehot}}^{M,K}(\Phi)$ in \eqref{eq: joint MSE lower bound 1}
is the MSE lower bound, $L=2^M$, $i_{L-1}=K-\sum_{l=0}^{L-2}i_l$,
\setcounter{equation}{39}
\begin{equation}\label{eq: joint mean distribution 1}
	q_{\Phi}^{i_0,\cdots,i_{L-1}}(\theta)=
	\prod_{l=0}^{L-2}C_{K-i_0\cdots-i_{l-1}}^{i_l}(\gamma_{\Phi,l}(\theta))^{i_l},
\end{equation}
and, 
$$\gamma_{\Phi,l}(\theta)=\mathbb{E}_X[[G_{\Phi}(X)]_l|\theta].$$
\begin{IEEEproof}
	See Appendix \ref{Proof of Proposiiton {Prop: Joint quantization 1}} for details.
\end{IEEEproof}
\end{Proposition}

By using \eqref{eq: joint mean distribution 1} and following the analysis in \eqref{minimize mmse 3}, the estimation MSE for $\theta$ at the FC is computed as \eqref{eq: joint MSE cost 1}.
From \eqref{eq: joint MSE lower bound 1} and \eqref{eq: joint MSE cost 1}, the optimization problems for the multi-bit one-hot quantizer and FC are derived as
\setcounter{equation}{41}
\begin{align}
&\underset{\Phi}{\min}\ \mathcal{L}_{\mathrm{onehot}}^{M,K}(\Phi),\\
&\underset{\{\Phi,\Psi\}}{\min}\ \mathcal{T}_{\mathrm{onehot}}^{M,K}(\Phi,\Psi).
\end{align}
Since the training for $\Phi$ and $\Psi$ based on deep learning is similar to the binary quantization scenario discussed in section \ref{sec: Sequential training of binary quantizer and FC}, we omit the details of training process.

\section{Simulation Results}
This section presents some simulations to validate the proposed method, and the corresponding simulation environment is specified as follows:
Initially, the samples of the desired parameter in both the data and test sets are obtained from an uniform distribution over the interval $[-1,1]$ \cite{6882252}. 
The observation samples from each sensor are contaminated by i.i.d. Gaussian noise with variance $\sigma^2$, and $f_X(\cdot|\theta)\sim \mathcal{N}(0,\sigma^2)$ denotes the conditional distribution of the observation $X$ with given desired parameter $\theta$.
The observation signal-to-noise ratio (SNR) in the simulation is defined as the power ratio of the desired parameter to the observation noise.

For the quantizer module, the number of FCLs in the probability controller DNN $G_{\Phi}$ is set as $N=3$, with the number of neurons in each FCL being $L_G =20$;
%$L_{g,O,1}=L_{g,I,2}=L_{g,O,2}=L_{g,I,3}=20$; 
for the estimator DNN $F_{\Psi}$ in FC module, the number of neurons in each FCL is set as $L_F=30$.
%$L_{f,O,1}=L_{f,I,2}=L_{f,O,2}=L_{f,I,3}=30$. 
$G_{\Phi}$ is trained by a data set of 50000 samples over 500 epochs, with the predetermined sensor quantity parameter in the training of $G_{\Phi}$ being $K_S$. Using the trained $G_{\Phi^*}$, $F_{\Psi}$ is then trained by the same data set over 500 epochs, with the predetermined sensor quantity parameter in the training of $F_{\Psi}$ being $K_F$.
Based on the pretest simulation, the maximum epoch number of 500 is deemed efficient enough to achieve the desired convergence of the model training under all scenarios.
Besides, to ensure consistency throughout the simulation, the default values for both $K_S$ and $K_F$ are set to 250. However, it is important to note that certain figures explicitly indicate the usage of different values for $K_S$ or $K_F$.
After the sequential training of $G_{\Phi}$ and $F_{\Psi}$, the entire trained system undergoes validation using an independent test set comprising 10,000 samples. The test set is generated separately from the training set, ensuring a rigorous evaluation of the system's performance and verifying that our model is not overfitted.
The whole training uses the ADMM optimizer\cite{o2017introduction} with a constant learning rate $l_r=0.001$. 
The implementation of the whole simulations is carried out using PyTorch 1.7.0\cite{o2017introduction} on an NVIDIA 2080 Super Max-Q GPU. Under the aforementioned simulation environment, the training duration for the complete system is approximately 2 hours for the binary model, 4 hours for the 2-bit parallel models, and 10 hours for the 2-bit one-hot models.
In the entire simulations, the DNN parameters are initialized by the widely adopted Kaiming Gaussian distribution method \cite{he2015delving}, which is the default setting in the Pytorch. Additionally, to ensure the reliability of the results, the simulation of each figure is repeated multiple times to verify their consistency.

For comparisons, we implement the sine-quantization-maximum-likelihood-fusion (SQMLF) method \cite{6882252}, which was proved to be optimal for the estimation of a parameter with uniform distribution under the ideal noiseless scenario. In addition, we also examine the Posterior Cramer-Rao Lower Bound (PCRLB) \cite{5184907}, being the minimum MSE to be achieved by any unbiased estimation method, as the performance limit of the distributed estimation with binary quantization.

\subsection{Asymptotically optimality and robustness of proposed algorithm}
\label{sec: Asymptotically optimality and robustness of proposed algorithm}

\begin{figure}[htbp]
	\vspace{0pt}
	\setlength{\belowcaptionskip}{-10pt}
	\centering
	\includegraphics[width=230pt]{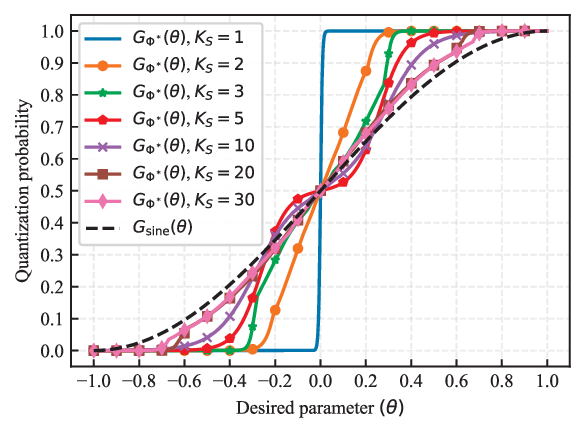}
	\caption{Probability controller $G_{\Phi}$ trained with different $K_S$.}
	\label{Fig: Trained quantizatoin probablity function 1}
\end{figure}
\vspace{10pt}
First, we study the asymptotic optimality and robustness of the proposed method to the number of sensors. Fig. \ref{Fig: Trained quantizatoin probablity function 1} plots the binary probability controller $G_{\Phi^*}$ trained with different $K_S$ under the noiseless observation scenario where local observations at sensors equal to the desired parameter. 
For the estimation of a parameter with uniform distribution under the noiseless scenario, the optimal probability controller function is proved to be $G_{\text{sine}}(\theta)=[1+\sin(\pi \theta/2)]/2$ used in the SQMLF method\cite{5184907}.
Therefore, the closer the trained $G_{\Phi^*}$ is to $G_{\text{sine}}$ under the noiseless scenario, the better quantitation performance and estimation performance are expected to be obtained. It is observed that with the increasing of $K_S$, the trained $G_{\Phi^*}$ gradually approaches to the optimal $G_{\text{sine}}$, which implies the asymptotic optimality of the proposed method.

\begin{figure}[htbp]
	\vspace{0pt}
	\setlength{\belowcaptionskip}{-10pt}
	\centering
	\includegraphics[width=230pt]{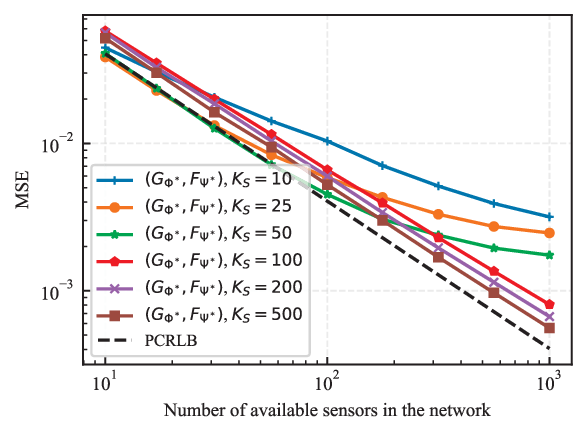}
	\caption{Estimation MSE vs. the number of available sensors in the network.}
	\label{Fig: MSE performance versus number of sensors for quantizer trianed with different K}
\end{figure}
Fig. \ref{Fig: MSE performance versus number of sensors for quantizer trianed with different K} plots the estimation MSE of the proposed method as a function of the number of available sensors in the network, with different $K_S$ being selected in the training of quantizer.
It is observed that selecting bigger $K_S$ in training improves the robustness of the proposed method to the variations of the number of sensors in the practical network. 
For instance, when $K_S=10$, although the estimation MSE is close to the PCRLB at the initial phase of the curves, it decreases more slowly and deviates further from the PCRLB as the number of sensors in test increases.
While with $K_S\geq 100$, the MSE decreases linearly with respect to the number of sensors and remains close to the PCRLB at the whole curves.
This validates the robustness of the proposed FC design with mean-fusion operation for accommodating various number of sensors in practice.

\begin{figure}[htbp]
	\vspace{0pt}
	\setlength{\belowcaptionskip}{-10pt}
	\centering
	\includegraphics[width=230pt]{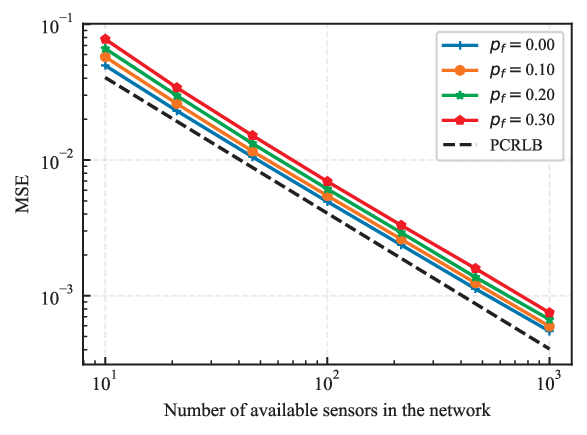}
	\caption{Robustness to the failure of sensors.}
	\label{Fig: sensor failure}
\end{figure}
To further evaluate the robustness of our method, we conducted additional simulations to assess its performance in scenarios with sudden sensor failures. We consider the situation where each sensor in the network randomly fails to transmit information to the FC with a probability denoted as $p_f$. Fig. \ref{Fig: sensor failure} illustrates the estimation MSE of our method under different values of $p_f$. The simulation result reveals a consistent linear decrease in the estimation MSE of our proposed method across varying $p_f$ values, as the number of sensors increases. Notably, the MSE remains close to the PCRLB under all conditions. 
This result provide empirical evidence that our method is robust against sudden sensor failures and possesses a high level of generalization to the network variation.

\begin{figure}[htbp]
	\normalsize
	\centering
	\subfloat[Probability controller DNN $G_{\Phi}$]{
		\includegraphics[width=225pt]{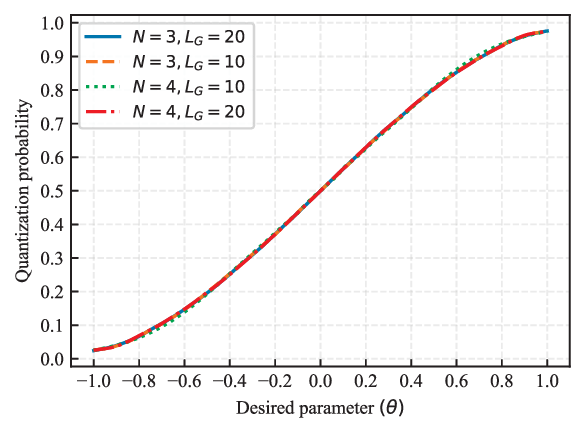}}\\
	\subfloat[Estimation MSE]{
		\includegraphics[width=230pt]{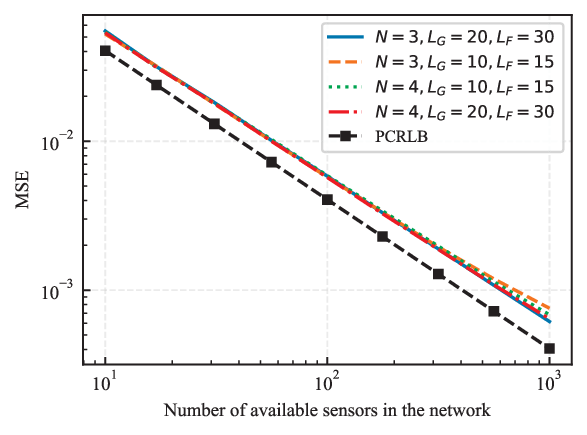}}
	\caption{Robustness to the neural network initialization.}
	\label{Fig: Robustness of the proposed method to the neural network initialization}
\end{figure}
To investigate the robustness of our method to the neural network initializations, we conducted the simulations of our proposed method with various initial setting parameters, including the number of layers $N$ in the DNNs, the number of neurons $L_G$ in each layer of $G_{\Phi}$ and the number of neurons $L_F$ in each layer of $F_{\Psi}$. Fig. \ref{Fig: Robustness of the proposed method to the neural network initialization}(a) plots the convergence results of the probability controller DNN $G_{\Phi}$ under different numbers of layers and neurons. It is observed that $G_{\Phi}$ converges to almost the same structure despite the variations in DNN setting parameters. Furthermore, Fig. \ref{Fig: Robustness of the proposed method to the neural network initialization}(b) portrays the estimation MSE of our method with different numbers of layers and neurons. 
Similarly, the estimation MSE curves of our method under diverse DNN setting parameters are almost the same, which verify the robustness of our method to the neural network initialization. Besides, it is observed that the minimum DNN initial configuration yielding comparable results for our proposed method is ``$N=3,L_G=10,L_F=15$". As depicted in Fig. 11 our manuscript, a slight increase in the MSE is observed when employing this minimum configuration compared to simulations with a larger number of neurons and layers. This observation implies that selecting a number of neurons and layers smaller than the aforementioned minimum values would result in a more significant increase in the MSE of our method.

\subsection{Noise suppression ability}
\begin{figure*}[htbp]
	\normalsize
	\centering
	\subfloat[SNR=-4dB]{
		\includegraphics[width=0.32\linewidth]{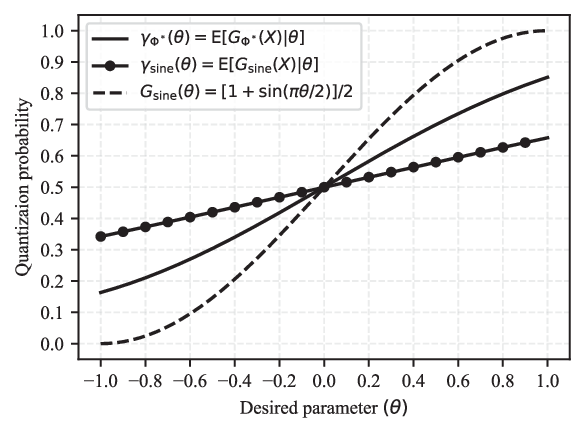}}
	\subfloat[SNR=0dB]{
		\includegraphics[width=0.32\linewidth]{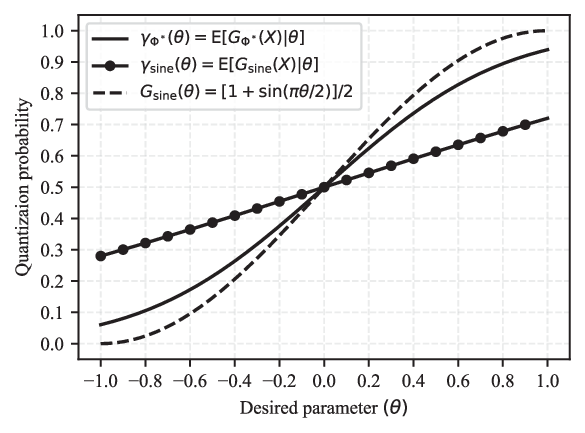}}
	\subfloat[SNR=4dB]{
		\includegraphics[width=0.32\linewidth]{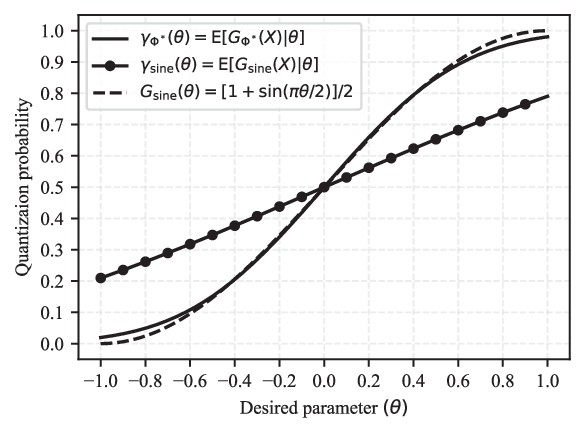}}\\
	\subfloat[SNR=8dB]{
		\includegraphics[width=0.32\linewidth]{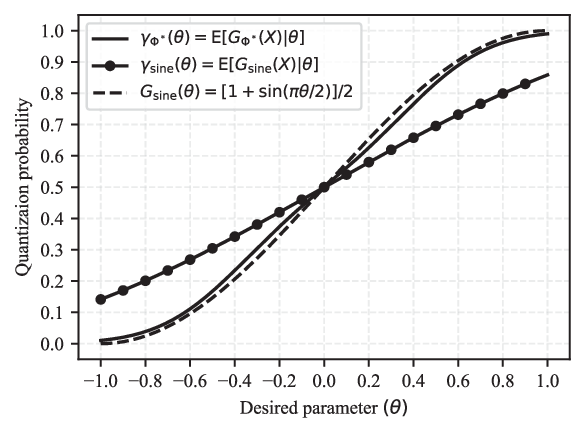}}
	\subfloat[SNR=12dB]{
		\includegraphics[width=0.32\linewidth]{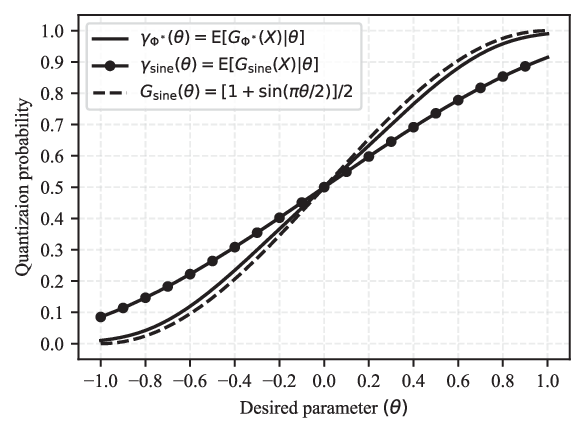}}
	\subfloat[SNR=16dB]{
		\includegraphics[width=0.32\linewidth]{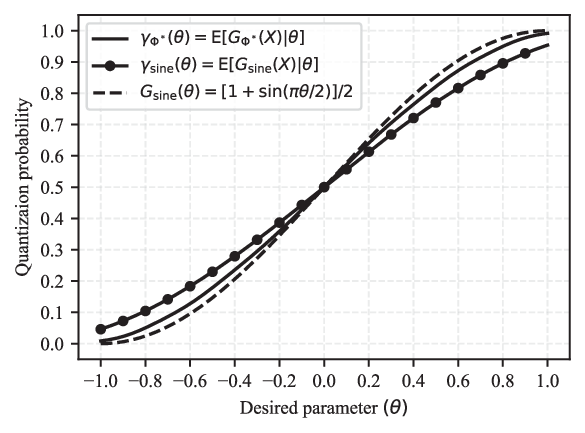}}
	\caption{Noisy quantization probability of the proposed and SQMLF methods under different SNRs.}
	\label{Fig: Noise suppression ability of observation noise 1}
\end{figure*}
In this section, we investigate the noise suppression ability and robustness of the proposed method to the observation noise, considering both the stable noisy scenario with SNR being constant during the training and test process, and the unstable scenario with fluctuating SNRs. Fig. \ref{Fig: Noise suppression ability of observation noise 1} plots the noisy quantization probability of the proposed method, i.e., $\gamma_{\Phi^*}(\theta)=\mathbb{E}_X[G_{\Phi^*}(X)|\theta]$ defined in \eqref{eq: gamma() 1}, under the stable noisy scenario.
Similar to the noiseless scenario, the optimal $\gamma_{\Phi^*}(\theta)$ should equal to $G_{\text{sine}}(\theta)$ \cite{5184907}. 
In particular, if $G_{\text{sine}}(X)$ is directly used as the probability controller at the sensor, its corresponding noisy quantization probability becomes $\gamma_{\text{sine}}(\theta)=\mathbb{E}_X[G_{\text{sine}}(X)|\theta]$
and is no longer optimal for the noisy scenario.
It is observed in Fig. \ref{Fig: Noise suppression ability of observation noise 1} that under differnt SNRs, $\gamma_{\Phi^*}(\theta)$ is closer to the optimal $G_{\text{sine}}(\theta)$ than $\gamma_{\text{sine}}(\theta)$. Besides, $\gamma_{\Phi^*}(\theta)$ is almost identical to $G_{\text{sine}}(\theta)$ when $\text{SNR}\ge 4\text{dB}$, whereas $\gamma_{\text{sine}}$ becomes almost identical to $G_{\text{sine}}(\theta)$ until $\text{SNR}$ reaches 16 dB.
This validates the superiority of the proposed method on the observation noise suppression.

\begin{figure}[htbp]
	\vspace{0pt}
	\setlength{\belowcaptionskip}{-10pt}
	\centering
	\subfloat[Proposed method]{\includegraphics[width=0.9\linewidth]{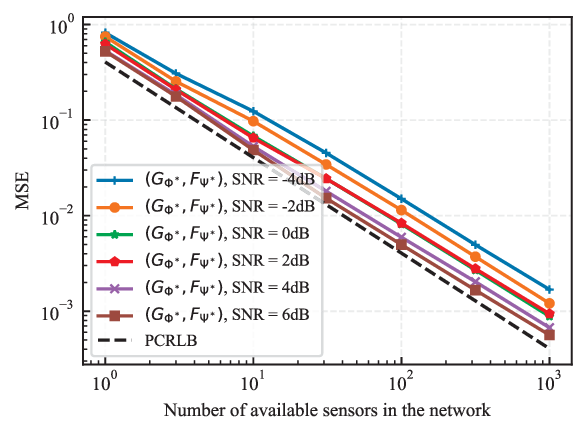}
	\label{Fig: Binary_noise MSE versus num_sensor with different SNR-proposed method}}\\
	\subfloat[SQMLF method]{\includegraphics[width=0.9\linewidth]{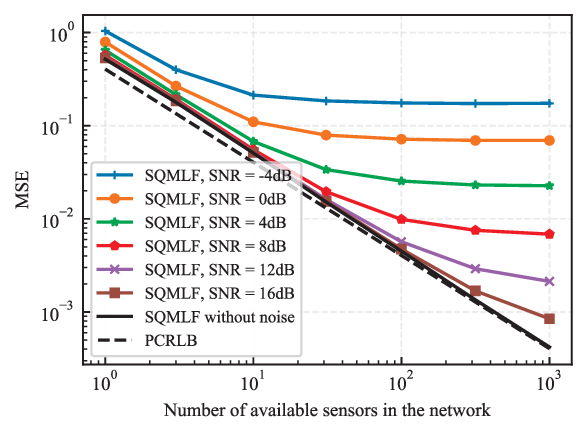}
	\label{Fig: Binary_noise MSE versus num_sensor with different SNR-SQMLF method}}
	\caption{Estimation MSE of the proposed and SQMLF methods under the stable observation scenario.}
	\label{Fig: Binary_noise MSE versus num_sensor with different SNR}
\end{figure}
Fig. \ref{Fig: Binary_noise MSE versus num_sensor with different SNR}(a) and Fig. \ref{Fig: Binary_noise MSE versus num_sensor with different SNR}(b) plot the estimation MSE of the proposed and SQMLF methods as a function of the number of sensors under the stable observation scenario.
It is observed that although the SQMLF method performs nearly optimally in the ideal noiseless scenario, its performance suffers severe degradation in the noisy scenario and its estimation MSE decreases more slowly with the increasing of the number of sensors. 
According to Fig. \ref{Fig: Noise suppression ability of observation noise 1}, this observation can be attributed to the fact that the deviation of the noisy quantization probability of the SQMLF method from the optimal structure becomes more pronounced as the SNR decreases, consequently resulting in a greater performance degradation.
In contrast, the estimation MSE of the proposed method under different SNRs remains close to the PCRLB and is linearly decreasing with respect to the number of sensors.

\begin{figure}[htbp]
	%\vspace{-10pt}
	%\setlength{\abovecaptionskip}{0pt}
	%\setlength{\belowcaptionskip}{0pt}
	\centering
	\includegraphics[width=230pt]{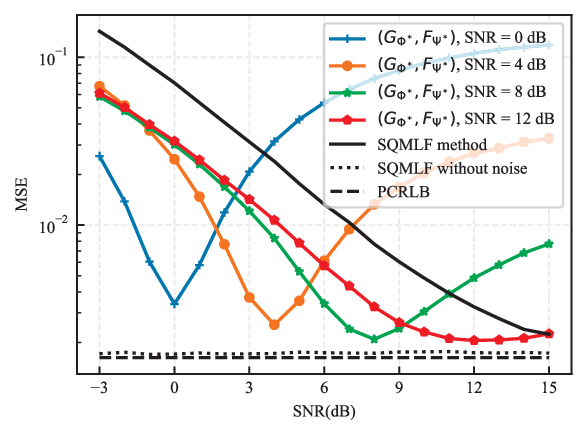}
	\caption{Estimation MSE of the proposed method under unstable observation scenario.}
	\label{Fig: Binary_noise MSE versus SNR with different trained_snr}
\end{figure}
Next, we evaluate the robustness of the proposed method in the presence of the unstable noisy observations with fluctuating SNRs. Fig. \ref{Fig: Binary_noise MSE versus SNR with different trained_snr} plots the estimation MSE as a function of the SNRs, with the number of sensors being set as 250. The estimation MSE of the proposed method trained under different SNRs are tested and compared with the SQMLF method. It is observed that with the increasing of the gap between the SNRs in test and training, the estimation MSE of the proposed method increases and presents to be a U-shape curve with respect to the SNRs. 
%This can be attributed to the quantization probability in the test being further from the optimal one in training as the SNR gap increases, leading to greater performance degradation. 
Consequently, in scenarios with very high SNRs, the performance of the proposed method trained under low SNRs is inferior to that of the SQMLF method. However, it is also observed that the proposed method trained under a higher SNR exhibits better robustness to the varying SNRs in test. For instance, the proposed method trained under SNR = 12 dB outperforms the SQMLF method for all SNRs tested. 

\begin{figure}[htbp]
	\vspace{0pt}
	\setlength{\belowcaptionskip}{-10pt}
	\centering
	\includegraphics[width=230pt]{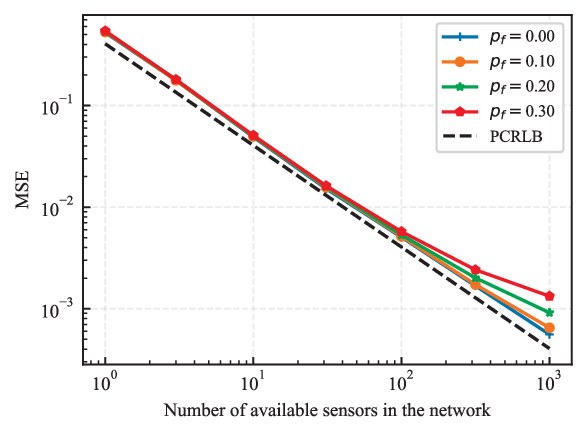}
	\caption{Robustness to a sudden spike in sensor noise,\\ with 4 times power spike being considered.}
	\label{Fig: sensor spike}
\end{figure}

In practical systems, it is reasonable to account for the possibility of a sudden spike in sensor noise. We consider the scenario that all sensors in the network randomly have a 4 times power spike in its observation noise, i.e., 6 dB SNR degradation at the sensors, with the probability denoted as $p_s$. Fig. \ref{Fig: sensor spike} plots the estimation MSE of our method under different values of $p_s$. It is observed that the estimation MSE of our proposed method remains close to the PCRLB under differnt number of sensors and values of $p_s$. Notably, there is only a slight increase in the MSE when the network has a huge number of sensors and a high sensor spike probability like $P_s = 0.3$. This result provides empirical evidence that our proposed method is robust against the sudden spike in sensor noise.

\subsection{Multi-bit quantization}

\begin{figure*}[htbp]
	\normalsize
	\centering
	\subfloat[$K_s=1$]{
		\includegraphics[width=0.32\linewidth]{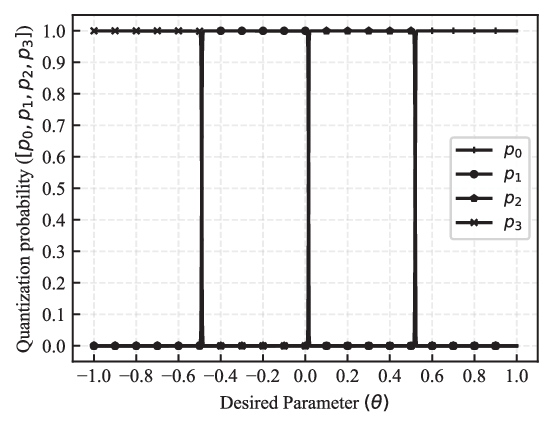}}
	\subfloat[$K_s=2$]{
		\includegraphics[width=0.32\linewidth]{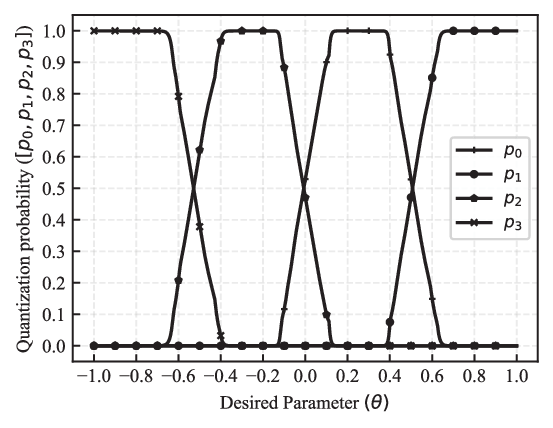}}
	\subfloat[$K_s=5$]{
		\includegraphics[width=0.32\linewidth]{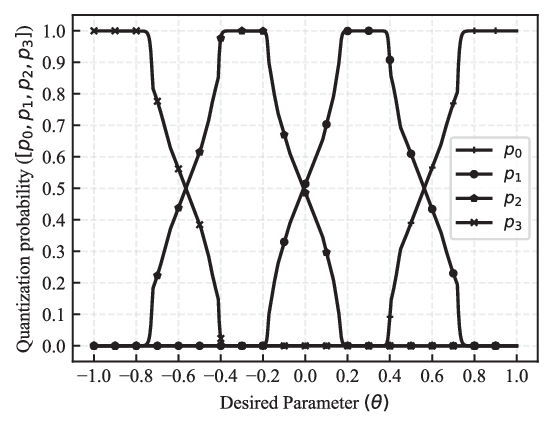}}\\
	\subfloat[$K_s=10$]{
		\includegraphics[width=0.32\linewidth]{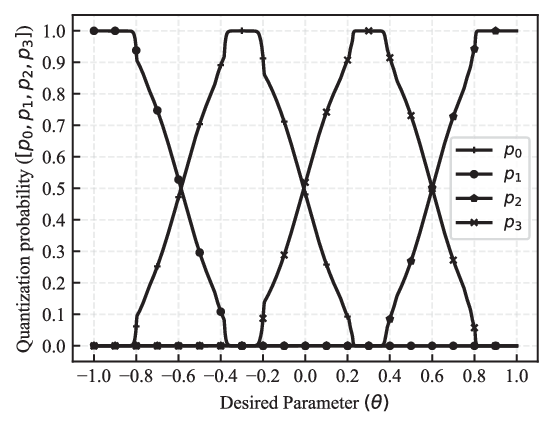}}
	\subfloat[$K_s=25$]{
		\includegraphics[width=0.32\linewidth]{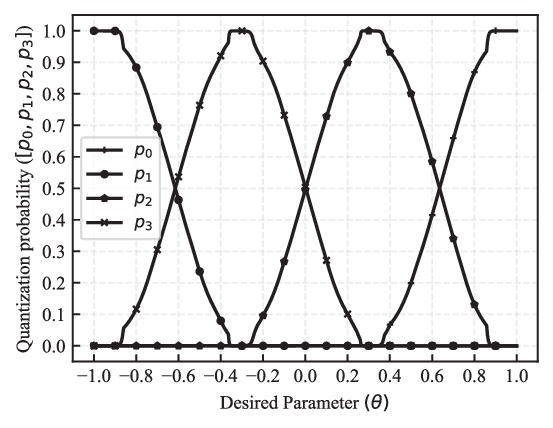}}
	\subfloat[$K_s=50$]{
		\includegraphics[width=0.32\linewidth]{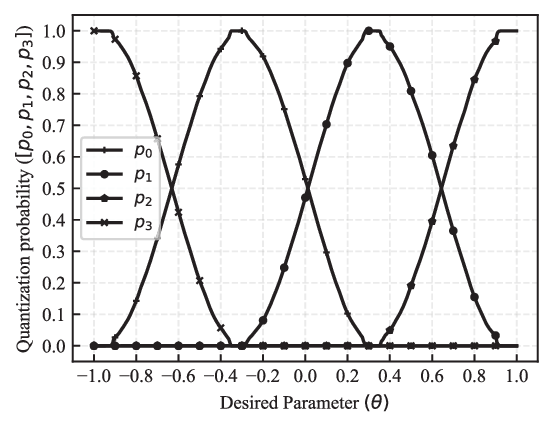}}
	\caption{2-bit one-hot probability controller $G_{\Phi}$ trained with different $K_S$.}
	\label{Fig: Asymptotically optimality of the multi-bit quantizer design.}
\end{figure*}
\begin{figure}[htbp]
	\vspace{0pt}
	\setlength{\belowcaptionskip}{-10pt}
	\centering
	\includegraphics[width=0.9\linewidth]{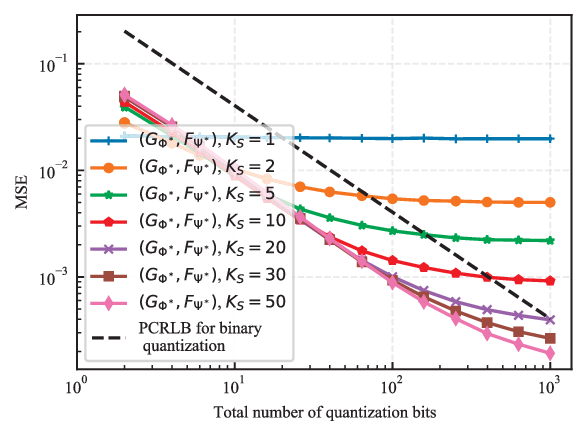}
	\caption{Estimation MSE of the 2-bit one-hot quantization vs. the total number of quantization bits, with $K_F=50$.}
	\label{Fig: MSE performance versus number total quantization bits for joint quantizer trianed with different K}
\end{figure}

\begin{figure}[htbp]
	%\vspace{-10pt}
	%\setlength{\abovecaptionskip}{0pt}
	%\setlength{\belowcaptionskip}{0pt}
	\centering
	\includegraphics[width=230pt]{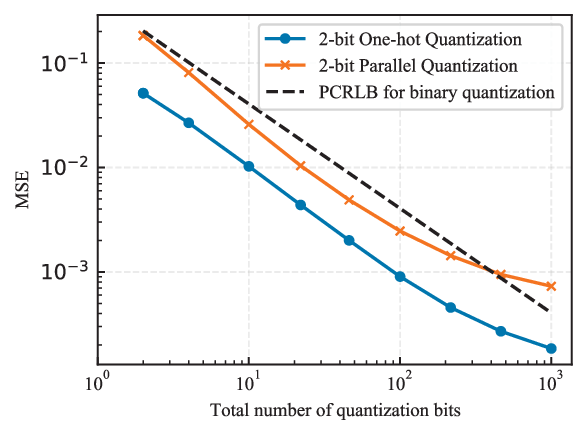}
	\caption{Estimation MSE of 2-bit one-hot quantization and 2-bit parallel quantization models, with $K_S=K_F=50$.}
	\label{Fig: 2-bit joint and parallel model MSE}
\end{figure}

Finally, we investigate the performance of the proposed multi-bit parallel quantization and one-hot quantization schemes, 
and we select the multi-bit quantization dimension $M=2$ for both the two models.
Fig. \ref{Fig: Asymptotically optimality of the multi-bit quantizer design.} plots the 2-bit one-hot probability controller $G_{\Phi^*}$ trained with different $K_S$ under the noiseless scenario. It is observed that as $K_S$ increases, the trained $G_{\Phi^*}$ converges to a stable structure with a stable quantization probability distribution on the desired parameter $\theta$. 
Fig. \ref{Fig: MSE performance versus number total quantization bits for joint quantizer trianed with different K} plots the estimation MSE of the 2-bit one-hot quantization scheme as a function of the total number of quantization bits from all sensors. It is observed that with the increasing of $K_S$, the performance of the 2-bit one-hot quantization becomes asymptotically convergent and optimal, outperforming the performance limit of any binary-quantization-based estimation scheme. The result verifies the asymptotic optimality of the proposed method.

Fig. \ref{Fig: 2-bit joint and parallel model MSE} plots the estimation MSE of both the 2-bit one-hot quantization and 2-bit parallel quantization schemes as a function of the total number of quantization bits from all sensors. The PCRLB of the binary quantization with respect to the number of quantization bits is plotted as a comparison. It is observed that the estimation performance of both the one-hot quantization and parallel quantitation schemes outperform the performance limit of any binary quantization scheme. 
Besides, it is observed that the one-hot quantization scheme exhibits better performance than the parallel one. This can be explained by the inclusion relationship between the two schemes:  
the set of any potential quantization probability distributions based on the parallel quantizer is a subset of those realizable by the one-hot quantizer.
Therefore, the performance ceiling of the one-hot quantization scheme is at least equivalent to or better than that of the parallel scheme.

\section{Conclusion}
In this paper, we propose a joint model and data driven method for the quantization-based distributed estimation problem. First, for sensors with binary quantization and conditionally i.i.d. observations, the MSE lower bound for the distributed estimation is derived, and a binary probabilistic quantizer is designed to minimize this lower bound. The optimality of the mean-fusion operation at the FC for the estimation with MMSE criterion is proved and a corresponding FC design is proposed. 
Considering the practical scenarios that the distributions of both the desired parameter and observation noise are unknown or only the noise distribution is known, a joint model and data driven method is proposed to train the probability controller in quantizer and the estimator in FC as DNNs.
By relaxing the binary quantization constraint at the sensors, the results are extended to the cases of multi-bit parallel quantization and one-hot quantization. Simulation results reveal that the proposed method outperforms state-of-the-art schemes under certain conditions.
%when the desired parameter and observation noise distributions are imperfectly known.

%%%%%%%%%%%%%%%%%%%%%%%%%%%%%%%%%%%%%%%%%%%%%%%%%%%%%%%%%%%%%%%%%%%%%
\appendices
\appendices
\section{Proof of Proposition \ref{MSE lower bound 1}}
\label{Proof of Proposition MSE lower bound 1}
\begin{figure*}[!b]
	\normalsize
	\newcounter{mytempeqncnt4}
	\setcounter{mytempeqncnt4}{\value{equation}}
	\hrulefill
	%%%
	\setcounter{equation}{46}
	\begin{equation}
		\label{MMSE lower bound 2}
		\begin{aligned}
			\mathbb{E}[|\theta-\hat{\theta}(\mathbf{u})|^2]
			\ge
			\mathbb{E}_{\theta}[\theta^2]-
			\sum_{k=0}^K\sum_{\mathbf{u}\in\mathcal{U}_K}
			\frac{\mathbb{E}_{\theta}^2\left[\theta p\left(\mathbf{u}\left|\theta\right.\right)\right]}
			{\mathbb{E}_{\theta}\left[ p\left(\mathbf{u}\left|\theta\right.\right)\right]}
			=
			\mathbb{E}_{\theta}[\theta^2]-
			\sum_{k=0}^KC_K^k
			\frac{\mathbb{E}_{\theta}^2\left[\theta (\gamma(\theta))^k(1-\gamma(\theta))^{K-k}\right]}
			{\mathbb{E}_{\theta}\left[ (\gamma(\theta))^k(1-\gamma(\theta))^{K-k}\right]}.
		\end{aligned}
	\end{equation} 
	%%%%
	\setcounter{equation}{49}
	\begin{equation}\label{eq: I_2(theta)}
		\begin{aligned}
			I_2(\theta)= \mathbb{E}_{\bar{u}}\left[-\frac{\partial^{2} \ln p(\bar{u} \mid \theta)}{\partial \theta^{2}}\right]
			=(\gamma'(\theta))^2\sum_{i=0}^KC_{K}^{i}(\gamma(\theta))^{i-2}(1-\gamma(\theta))^{K-i-2}\left[i(1-\gamma(\theta))-(K-i)\gamma(\theta)\right]^2
			%=K\frac{(\gamma'(\theta))^2}{\gamma(\theta)(1-\gamma(\theta)}
			= I_1(\theta).
		\end{aligned}
	\end{equation} 
	%%%%
	\begin{equation} 
		\label{MMSE lower bound 3}
		\begin{aligned}
			\mathbb{E}[|\theta-\hat{\theta}(\bar{u})|^2]
			\ge \mathbb{E}_{\theta,\bar{u}}
			[|\theta-\hat{\theta}_{\text{MMSE}}(\bar{u})|^2]
			=\mathbb{E}_{\theta}[\theta^2]-
			\sum_{k=0}^KC_K^k
			\frac{\mathbb{E}_{\theta}^2\left[\theta (\gamma(\theta))^k(1-\gamma(\theta))^{K-k}\right]}
			{\mathbb{E}_{\theta}\left[ (\gamma(\theta))^k(1-\gamma(\theta))^{K-k}\right]}.
		\end{aligned}
	\end{equation}
	%%%%
	\setcounter{equation}{\value{mytempeqncnt4}}
	%\hrulefill
	\vspace*{4pt}
\end{figure*}
The FC utilizes the quantized data $\mathbf{u}=[u_1\cdots,u_K]^T$ from all sensors to estimate the desired parameter $\theta$ as $\hat{\theta}(\mathbf{u})$, and the estimator $\hat{\theta}(\cdot)$ determines the estimation MSE for $\theta$. According to \cite{schonhoff2006detection}, the MMSE estimator for the estimation of $\theta$ with given $\mathbf{u}$ is derived as 
\begin{equation}\label{MMSE estimator of theta given vector u}
	\begin{aligned}
		\hat{\theta}_{\text{MMSE}}\left({\mathbf{u}}\right)
		=
		\mathbb{E}_{\theta,\mathbf{u}}\left[\theta\left|\mathbf{u}\right.\right]
		=
		\frac{\mathbb{E}_{\theta}\left[\theta p\left(\mathbf{u}\left|\theta\right.\right)\right]}
		{\mathbb{E}_{\theta}\left[ p\left(\mathbf{u}\left|\theta\right.\right)\right]},
	\end{aligned}
\end{equation}
and the achievable MSE in estimating $\theta$ using $\mathbf{u}$ is lower bounded by
\begin{equation}
	\label{MMSE lower bound 1}
	\begin{aligned}
		\mathbb{E}[|\theta-\hat{\theta}(\mathbf{u})|^2]
		\ge&\ 
		\mathbb{E}_{\theta,\mathbf{u}}
		[|\theta-\hat{\theta}_{\text{MMSE}}(\mathbf{u})|^2]\\
		=&\
		\mathbb{E}_{\theta}[\theta^2]-\sum_{\mathbf{u}\in\mathcal{U}}
		\frac{\mathbb{E}_{\theta}^2\left[\theta p\left(\mathbf{u}\left|\theta\right.\right)\right]}
		{\mathbb{E}_{\theta}\left[ p\left(\mathbf{u}\left|\theta\right.\right)\right]},
	\end{aligned}
\end{equation}
where $\mathcal{U}=\{0,1\}^K$ is the set of all possible results for $\mathbf{u}$.

When the quantized data from all sensors are conditionally i.i.d. with given $\theta$, the above results can be further simplified. Define $\{\mathcal{U}_k\}_{k=0}^K$ as the sequence of non-overlapping subsets of $\mathcal{U}$, with  $\mathcal{U}_k=\{\mathbf{u}\in\mathcal{U}|\frac{1}{K}\sum_{i=1}^K u_i=k/K\}$. It's intuitively to see that $\mathcal{U}=\cup_{k=0}^K\mathcal{U}_k$ and $|\mathcal{U}_k|=C_K^k$. Besides, any $\mathbf{u}$ belonging to set $\mathcal{U}_k$ have the identical conditional probability with given $\theta$, i.e.,
\begin{equation}\label{bernoulli distribution 1}
	\begin{aligned}		 
		p(\mathbf{u}|\theta)
		=(\gamma(\theta))^k(1-\gamma(\theta))^{K-k},
	\end{aligned}
\end{equation}   
$\forall$ $\mathbf{u}\in \mathcal{U}_k$, where $\gamma(\theta)=p(u=1|\theta)$ denotes the conditional probability of any quantized data being '1' with given $\theta$. By subsisting \eqref{bernoulli distribution 1} into \eqref{MMSE lower bound 1} and utilizing $|\mathcal{U}_k|=C_K^k$, the MSE lower bound in \eqref{MMSE lower bound 1} is rewritten as \eqref{MMSE lower bound 2}.
This completes the proof.

\section{Proof of Proposition \ref{Mean Value Fisher Information 1}}
\label{Proof of Proposition Mean Value Fisher Information 1}
According to \cite{5184907}, the Fisher Information of $\mathbf{u}$ with given $\theta$ is derived as
\setcounter{equation}{47}
\begin{equation}
	\begin{aligned}
		I_1(\theta)
		&=\mathbb{E}_{\mathbf{u}}\left[-\frac{\partial^{2} \ln p(\mathbf{u} \mid \theta)}{\partial \theta^{2}}\right]\\
		&=K \frac{(\gamma'(\theta))^2}{\gamma(\theta)(1-\gamma(\theta))},
	\end{aligned}
\end{equation}	
where $\gamma(\theta)$ is defined in \eqref{eq: gamma() 1}. 
Based on \eqref{bernoulli distribution 1}, the conditional probability distribution for the average  $\bar{u}$ of all quantized data with given $\theta$ is computed as
\begin{equation}\label{eq: bernoulli binomial distribution 2}
	\begin{aligned}
		p\left(\bar{u}=\frac{k}{K}\bigg|\theta\right)
		=\ p(\mathbf{u}\in\mathcal{U}_k|\theta)
		=\ C_K^k\left(\gamma(\theta)\right)^k
		\left(1-\gamma(\theta)\right)^{K-k},
	\end{aligned}
\end{equation}
for $k=0,1,\cdots,K$. Then, by using \eqref{eq: bernoulli binomial distribution 2}, the Fisher Information of $\bar{u}$ with given $\theta$ is derived as \eqref{eq: I_2(theta)}.			

Similar to \eqref{MMSE estimator of theta given vector u}, the MMSE estimator for estimating $\theta$ by using $\bar{u}$ is obtained as $\hat{\theta}_{\text{MMSE}}(\bar{u})=\mathbb{E}_{\theta}\left[\theta p\left(\bar{u}\left|\theta\right.\right)\right]/\mathbb{E}_{\theta}\left[ p\left(\bar{u}\left|\theta\right.\right)\right]$. The achievable MSE lower bound for the estimation of $\theta$ with $\bar{u}$ is derived as \eqref{MMSE lower bound 3}, and is identical to the achievable MSE lower bound for estimating $\theta$ with $\mathbf{u}$ as derived in \eqref{eq: mse lower bound 1}. \eqref{eq: I_2(theta)} and \eqref{MMSE lower bound 3} complete the proof.

\section{Proof of Proposition \ref{Prop: Paraller quantization 1}}
\label{Proof of Proposiiton {Prop: Paraller quantization 1}}
\begin{figure*}[!b]
	\normalsize
	\newcounter{mytempeqncnt5}
	\setcounter{mytempeqncnt5}{\value{equation}}
	\hrulefill
	%%%
	\setcounter{equation}{56}
	\begin{equation}\label{bernoulli distribution 6}
		\begin{aligned}
			p\left(\bar{\mathbf{u}}=\frac{1}{K}[i_1,\cdots,i_M]^T\bigg|\theta\right)
			=p\left(\mathbf{U}\in\mathcal{U}_{i_1,\cdots,i_M}\bigg|\theta\right)
			=\prod_{m=1}^MC_K^{i_m}(\gamma_{\phi_m}(\theta))^{i_m}(1-\gamma_{\phi_m}(\theta))^{K-i_m},
		\end{aligned}
	\end{equation}
	%%%%
	\begin{equation}
		\label{MMSE lower bound 4}
		\begin{aligned}
			\mathbb{E}[|\theta-\hat{\theta}(\mathbf{U})|^2]
			\ge
			\mathbb{E}_{\theta,\mathbf{U}}
			[|\theta-\hat{\theta}_{\text{MMSE}}(\mathbf{U})|^2]
			=\mathbb{E}[\theta^2]-
			\sum_{i_1=0}^{K}
			\cdots
			\sum_{i_{M}=0}^{K}
			\frac{\mathbb{E}_{\theta}^2\left[\theta p\left(\mathbf{U}\in\mathcal{U}_{i_1,\cdots,i_M}\bigg|\theta\right)\right]}
			{\mathbb{E}_\theta\left[p\left(\mathbf{U}\in\mathcal{U}_{i_1,\cdots,i_M}\bigg|\theta\right)\right]},
		\end{aligned}
	\end{equation} 
	%%%%
	\begin{equation}
		\label{MMSE lower bound 5}
		\begin{aligned}
			\mathbb{E}[|\theta-\hat{\theta}(\bar{\mathbf{u}})|^2]
			\ge
			\mathbb{E}_{\theta,\bar{\mathbf{u}}}
			[|\theta-\hat{\theta}_{\text{MMSE}}(\bar{\mathbf{u}})|^2]
			=\mathbb{E}[\theta^2]-
			\sum_{i_1=0}^K\cdots\sum_{i_M=0}^K
			\frac{\mathbb{E}_{\theta}^2\left[\theta p\left(\bar{\mathbf{u}}=\frac{1}{K}[i_1,\cdots,i_M]^T\left|\theta\right.\right)\right]}
			{\mathbb{E}_{\theta}\left[ p\left(\bar{\mathbf{u}}=\frac{1}{K}[i_1,\cdots,i_M]^T\left|\theta\right.\right)\right]},
		\end{aligned}
	\end{equation} 
	%%%%
	\setcounter{equation}{62}
	\begin{equation}
		\label{MMSE lower bound 6}
		\begin{aligned}
			\mathbb{E}[|\theta-\hat{\theta}(\mathbf{U})|^2]
			\ge
			\mathbb{E}_{\theta,\mathbf{V}}
			[|\theta-\hat{\theta}_{\text{MMSE}}(\mathbf{V})|^2]
			=\mathbb{E}[\theta^2]-
			\sum_{i_1=0}^{K}
			\cdots
			\sum_{i_k=0}^{L-i_1-\cdots-i_{k-1}}
			\cdots
			\sum_{i_{L-1}=0}^{K-i_0-\cdots-i_{L-2}}
			\frac{\mathbb{E}^2\left[\theta p\left(\mathbf{V}\in\mathcal{V}_{i_0,\cdots,i_{L-1}}\bigg|\theta\right)\right]}
			{\mathbb{E}\left[p\left(\mathbf{V}\in\mathcal{V}_{i_0,\cdots,i_{L-1}}\bigg|\theta\right)\right]},
		\end{aligned}
	\end{equation} 
	%%%
	\begin{equation}
		\label{MMSE lower bound 7}
		\begin{aligned}
			\mathbb{E}[|\theta-\hat{\theta}(\bar{\mathbf{v}})|^2]
			\ge
			\mathbb{E}_{\theta,\bar{\mathbf{v}}}
			[|\theta-\hat{\theta}_{\text{MMSE}}(\bar{\mathbf{v}})|^2]
			=\mathbb{E}[\theta^2]-
			\sum_{i_0=0}^{K}
			\cdots
			\sum_{i_k=0}^{L-i_1-\cdots-i_{k-1}}
			\cdots
			\sum_{i_{L-1}=0}^{K-i_1-\cdots-i_{L-2}}
			\frac{\mathbb{E}_{\theta}^2\left[\theta p\left(\bar{\mathbf{v}}=\frac{1}{K}[i_0,\cdots,i_{L-1}]^T\bigg|\theta\right)\right]}
			{\mathbb{E}_\theta\left[p\left(\bar{\mathbf{v}}=\frac{1}{K}[i_0,\cdots,i_{L-1}]^T\bigg|\theta\right)\right]},
		\end{aligned}
	\end{equation} 
	%%%
	\setcounter{equation}{\value{mytempeqncnt5}}
	%\hrulefill
	\vspace*{4pt}
\end{figure*}
According to Proposition \ref{Prop: Conditional i.i.d. Proof 1}, when all sensors adopt the identical $M$-bit parallel quantizer with design parameter $\Phi$, the quantized data vectors $\mathbf{u}_1, \mathbf{u}_2, \cdots, \mathbf{u}_K\in\{0,1\}^M$ from all $K$ sensors are conditionally i.i.d., i.e.,
\setcounter{equation}{51}
\begin{align}\label{eq: conditionally i.i.d. 1}
	p(\mathbf{u}_i=\mathbf{u}|\theta)&=p(\mathbf{u}_j=\mathbf{u}|\theta),\\
	\label{eq: conditionally i.i.d. 2}
	p(\mathbf{u}_1,\cdots,\mathbf{u}_K|\theta)&=\prod_{k=1}^Kp(\mathbf{u}_k|\theta),
\end{align}
$\forall i,j\in \{1,\cdots,K\}, \forall\mathbf{u}\in\{0,1\}^M$. 
Then, it can be inferred from the parallel quantizer configuration in Fig. \ref{noise_case_multi_bit 1} that all entries of a quantized vector are conditionally independent with given $\theta$, i.e.,
\begin{equation}
	\begin{aligned}\label{eq: conditionally i.i.d. 3}
		p(\mathbf{u}_k|\theta)
		=&\prod_{m=1}^Mp([\mathbf{u}_k]_m|\theta),
	\end{aligned}
\end{equation}
for $k=1,2,\cdots,K$.
Based on \eqref{eq: conditionally i.i.d. 1}, \eqref{eq: conditionally i.i.d. 2} and \eqref{eq: conditionally i.i.d. 3}, it can be concluded that all entries in the quantized data matrix $\mathbf{U}=[\mathbf{u}_1^T,\cdots,\mathbf{u}_K^T]^T$ are conditionally independent with given $\theta$, i.e.,
\begin{equation}\label{eq: conditionally distirubted for all bits 1}
	p(\mathbf{U}|\theta)=\prod_{k=1}^Kp(\mathbf{u}_k|\theta)=\prod_{k=1}^K\prod_{m=1}^Mp([\mathbf{u}_k]_m|\theta).
\end{equation}
Define $\mathcal{U}=\{0,1\}^{K\times M}$ as the set of all possible results for $\mathbf{U}$, along with
a sequence of its non-overlapping subsets as $\mathcal{U}_{i_1,\cdots,i_M}=\{\mathbf{U}\in\mathcal{U}|\frac{1}{K}\sum_{i=1}^K[\mathbf{U}]_{i,m}=\frac{1}{K}\sum_{i=1}^K[\mathbf{u}_i]_{m}=\frac{i_m}{K} \ \text{for}\ m=1,\cdots,M\}$ for $i_1,\cdots,i_M\in\{0,\cdots,K\}$. It's intuitively to see that $\mathcal{U}=\cup_{i_1=0}^K\cdots\cup_{i_M=0}^K\mathcal{U}_{i_1,\cdots,i_M}$ and $|\mathcal{U}_{i_1,\cdots,i_M}|=\prod_{m=1}^MC_K^{i_m}$. From \eqref{eq: conditionally distirubted for all bits 1}, we have that if $\mathbf{U}\in\mathcal{U}_{i_1,\cdots,i_M}$, then
\begin{equation}\label{bernoulli distribution 5}
	\begin{aligned}		 
		p(\mathbf{U}|\theta)
		=&\prod_{m=1}^Mp\left(\frac{1}{K}\sum\nolimits_{i=1}^K[\mathbf{u}_i]_{m}=\frac{i_m}{K}\bigg|\theta\right)\\
		=&\prod_{m=1}^M(\gamma_{\phi_m}(\theta))^{i_m}(1-\gamma_{\phi_m}(\theta))^{K-i_m},
	\end{aligned}
\end{equation}   
where
$$\gamma_{\phi_m}(\theta)=p_{\phi_m}([\mathbf{u}_k]_m=1|\theta), k=1,\cdots,K.$$

Define $\bar{\mathbf{u}}=\frac{1}{K}\sum_{k=1}^K\mathbf{u}_k\in\{0,1/K,\cdots,1\}^M$ as the mean vector of the quantized data matrix. Based on \eqref{eq: conditionally distirubted for all bits 1} and \eqref{bernoulli distribution 5}, the conditional probability distribution of $\bar{\mathbf{u}}$ with given $\theta$ is derived as \eqref{bernoulli distribution 6},
$\forall\ i_1,\cdots,i_M \in\{0,1,\cdots,K\}$.
By utilizing \eqref{bernoulli distribution 5} and $|\mathcal{U}_{i_1,\cdots,i_m}|=\prod_{m=1}^MC_K^{i_M}$, the achievable MSE lower bound for estimating $\theta$ using $\mathbf{U}$ is derived as \eqref{MMSE lower bound 4},
where $\hat{\theta}_{\text{MMSE}}(\mathbf{U})=\mathbb{E}_{\theta}\left[\theta p\left(\mathbf{U}\left|\theta\right.\right)\right]/\mathbb{E}_{\theta}\left[ p\left(\mathbf{U}\left|\theta\right.\right)\right]$
is the MMSE estimator for the estimation of $\theta$ with $\mathbf{U}$.
From \eqref{bernoulli distribution 6}, the achievable MSE lower bound for estimating $\theta$ with $\bar{\mathbf{u}}$ is derived as \eqref{MMSE lower bound 5},
where $\hat{\theta}_{\text{MMSE}}(\bar{\mathbf{u}})=\mathbb{E}_{\theta}\left[\theta p\left(\bar{\mathbf{u}}\left|\theta\right.\right)\right]/\mathbb{E}_{\theta}\left[ p\left(\bar{\mathbf{u}}\left|\theta\right.\right)\right]$
is the corresponding MMSE estimator. By subsisting \eqref{bernoulli distribution 6} into \eqref{MMSE lower bound 4} and  \eqref{MMSE lower bound 5}, we complete the proof..

\section{Proof of Proposition \ref{Prop: Joint quantization 1}}
\label{Proof of Proposiiton {Prop: Joint quantization 1}}
Since the one-hot representation of a binary information is invertible, estimating the desired parameter $\theta$ with either the original quantized matrix $\mathbf{U}$ or its one-hot representation $\mathbf{V}$ achieves the identical estimation MSE, i.e.,
\setcounter{equation}{59}
\begin{equation}\label{eq: ivariance of U and V 1}
	\mathrm{E}[|\theta-\hat{\theta}(\mathbf{U})|^2]=
	\mathrm{E}[|\theta-\hat{\theta}(\mathbf{V})|^2].
\end{equation}
Similar to the analysis in \eqref{eq: conditionally i.i.d. 1} and \eqref{eq: conditionally i.i.d. 2}, the quantized vectors $\mathbf{u}_1, \cdots, \mathbf{u}_K$ are conditionally i.i.d. with given $\theta$, and so as their one-hot representation $\mathbf{v}_1, \cdots, \mathbf{v}_K$. Define
$$\mathcal{V}=\left\{\mathbf{V}\in \{0,1\}^{K\times L}\left|\sum_{l=0}^{L-1}[\mathbf{V}]_{k,l}=1, \forall \ k\in [K]\right.\right\}$$ 
as the set of all possible results for $\mathbf{V}$, with $L=2^M$. 
Define a sequence of non-overlapping subsets of $\mathcal{V}$ as $$\mathcal{V}_{i_0,\cdots,i_{L-1}}=\left\{\mathbf{V}\in\mathcal{V}\left|\sum_{k=1}^K[\mathbf{V}]_{k,l}=i_l\ \text{for}\ l=0,\cdots,L-1\right.\right\},$$
for all positive integers $i_0,\cdots,i_{L-1}$ satisfying  $\sum_{l=0}^{L-1}i_l=K$. It's intuitively to see that $\mathcal{V}=\cup_{i_0,\cdots,i_{L-1}}\mathcal{V}_{i_0,\cdots,i_{L-1}}$ and $|\mathcal{V}_{i_0,\cdots,i_{L-1}}|=\prod\nolimits_{l=0}^{L-1}C_{K-i_0-\cdots-i_{l-1}}^{i_l}.$

The conditional probability distribution of $\mathbf{V}$  with given $\theta$ is derived as
\begin{equation}\label{bernoulli distribution 7}
	p(\mathbf{V}|\theta)=\prod\nolimits_{l=0}^{L-1} (\gamma_{\Phi,l}(\theta))^{i_l},
\end{equation}
$\forall \mathbf{V}\in\mathcal{V}_{i_0,\cdots,i_{L-1}}$., where
$\gamma_{\Phi,l}(\theta)=\mathbb{E}_X[[G_{\Phi}(X)]_l|\theta].$
Define $\bar{\mathbf{v}}=\frac{1}{K}\sum_{k=1}^K\mathbf{v}_k$ as the mean-one-hot-vector of $\mathbf{U}$. Based on \eqref{bernoulli distribution 7}, the conditional probability distribution of $\bar{\mathbf{v}}$ with given $\theta$ is derived as
\begin{equation}\label{bernoulli distribution 8}
	\begin{aligned}
		p\left(\bar{\mathbf{v}}=\frac{1}{K}[i_0,\cdots,i_{L-1}]^T\bigg|\theta\right)
		=&p\left(\mathbf{V}\in\mathcal{V}_{i_0,\cdots,i_{L-1}}\bigg|\theta\right)\\
		=&\prod_{l=0}^{L-1}C_{K-i_0-\cdots-i_{l-1}}^{i_l}(\gamma_{\Phi,l}(\theta))^{i_l},
	\end{aligned}
\end{equation}
for all positive integers $i_0,\cdots,i_{L-1}$ satisfying  $\sum_{l=0}^{L-1}i_l=K$.

By utilizing \eqref{eq: ivariance of U and V 1} and \eqref{bernoulli distribution 7}, the achievable MSE lower bound for estimating $\theta$ with $\mathbf{U}$ is derived as \eqref{MMSE lower bound 6},
where $\hat{\theta}_{\text{MMSE}}(\mathbf{V})=\mathbb{E}_{\theta}\left[\theta p\left(\mathbf{V}\left|\theta\right.\right)\right]/\mathbb{E}_{\theta}\left[ p\left(\mathbf{V}\left|\theta\right.\right)\right]$
is the MMSE estimator for estimating $\theta$ using $\mathbf{V}$. The achievable MSE lower bound for estimating $\theta$ using $\bar{\mathbf{v}}$ is derived as \eqref{MMSE lower bound 7},
where $\hat{\theta}_{\text{MMSE}}(\bar{\mathbf{v}})=
\mathbb{E}_{\theta}\left[\theta p\left(\bar{\mathbf{v}}\left|\theta\right.\right)\right]/\mathbb{E}_{\theta}\left[ p\left(\bar{\mathbf{v}}\left|\theta\right.\right)\right]$
is the corresponding MMSE estimator. By subsisting \eqref{bernoulli distribution 8} into \eqref{MMSE lower bound 6} and  \eqref{MMSE lower bound 7}, we complete the proof.
	% you can choose not to have a title for an appendix
	% if you want by leaving the argument blank
	% \section{Computation of $\lambda_I$ and $\lambda_S$}  
	% \label{Power of the residual phase noise}
	
	% use section* for acknowledgment
	%\section*{Acknowledgment}
	
	%The authors would like to thank...
	
	% Can use something like this to put references on a page
	% by themselves when using endfloat and the captionsoff option.
	\ifCLASSOPTIONcaptionsoff
	\newpage
	\fi
	% trigger a \newpage just before the given reference
	% number - used to balance the columns on the last page
	% adjust value as needed - may need to be readjusted if
	% the document is modified later
	%\IEEEtriggeratref{8}
	% The "triggered" command can be changed if desired:
	%\IEEEtriggercmd{\enlargethispage{-5in}}

	% references section
	% can use a bibliography generated by BibTeX as a .bbl file
	% BibTeX documentation can be easily obtained at:
	% http://mirror.ctan.org/biblio/bibtex/contrib/doc/
	%\bibliographystyle{IEEEtran}
	% argument is your BibTeX string definitions and bibliography database(s)
	%\bibliography{IEEEabrv,../bib/paper}
	%
	% <OR> manually copy in the resultant .bbl file
	% set second argument of \begin to the number of references
	% (used to reserve space for the reference number labels box)
	\bibliographystyle{IEEEtran}
	\bibliography{citation}
	
	% biography section
	% 
	% If you have an EPS/PDF photo (graphicx package needed) extra braces are
	% needed around the contents of the optional argument to biography to prevent
	% the LaTeX parser from getting confused when it sees the complicated
	% \includegraphics command within an optional argument. (You could create
	% your own custom macro containing the \includegraphics command to make things
	% simpler here.)
	%\begin{IEEEbiography}[{\includegraphics[width=1in,height=1.25in,clip,keepaspectratio]{mshell}}]{Michael Shell}
	% or if you just want to reserve a space for a photo:
	
	\begin{IEEEbiography}[{\includegraphics[width=1in,height=1.25in,clip,keepaspectratio]{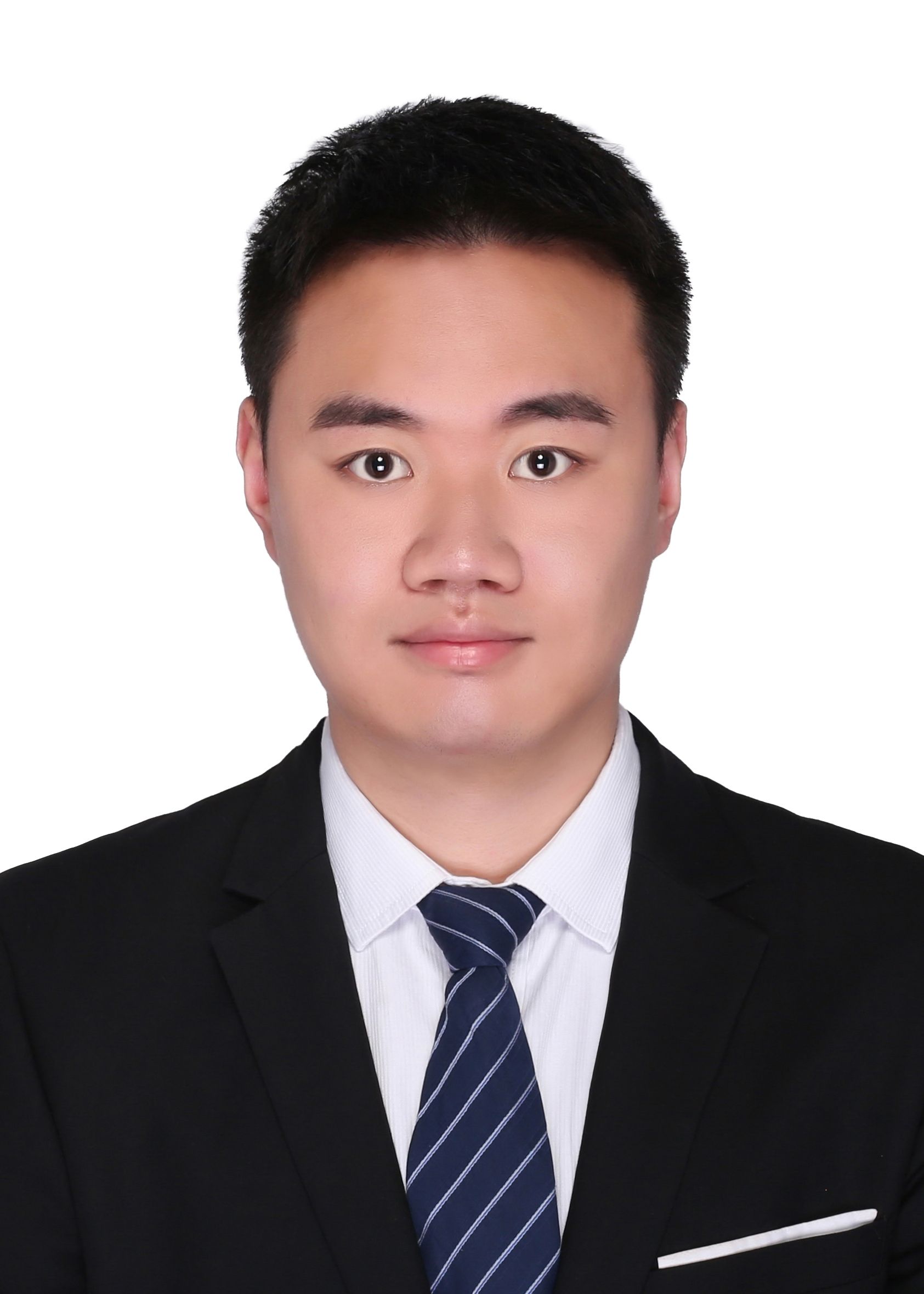}}]{Meng He}
		(Member, IEEE) received the B.E. degree in communication engineering from University of Electronic Science and Technology of China, Chengdu, China, in 2017, and the Ph.D. degree in computer and information engineering from The Chinese University of Hong Kong, Shenzhen, China, in 2023. He was a TPC member for IEEE GLOBECOM 2019-2022. He has been serving as a reviewer for IEEE TRANSACTIONS ON WIRELESS COMMUNICATIONS, and Journal of Communications and Information Networks. His current research interests include full-duplex communications, distributed estimation in wireless sensor networks and deep learning.
	\end{IEEEbiography}
	
	\begin{IEEEbiography}[{\includegraphics[width=1in,height=1.25in,clip,keepaspectratio]{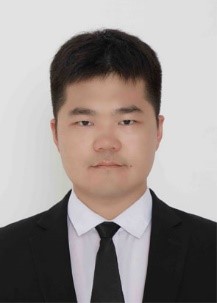}}]{Ran Li}
		(Member, IEEE) received the B.E. degree in communication engineering from University of Electronic Science and Technology of China, Chengdu, China, in 2017, and the Ph.D. degree in computer and information engineering from The Chinese University of Hong Kong, Shenzhen, China, in 2023. His current research interests include reinforcement learning and resource allocation in wireless networks.
	\end{IEEEbiography}
	
	\begin{IEEEbiography}[{\includegraphics[width=1in,height=1.25in,clip,keepaspectratio]{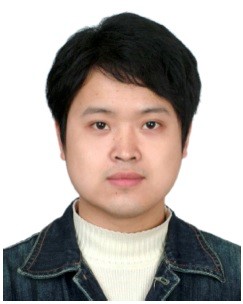}}]{Chuan Huang}
	(S’09–M’13) received his Ph.D. in Electrical Engineering from Texas A\&M University, College Station, Texas, USA, in 2012. From
	August 2012 to July 2014, he was a Research Associate with Princeton University, Princeton, NJ, USA, and then a Research Assistant Professor with
	Arizona State University, Tempe, AZ, USA. He is currently an Associate Professor with The Chinese University of Hong Kong, Shenzhen. His current
	research interests include wireless communications and signal processing. He has been serving as an Editor for IEEE TRANSACTIONS ON WIRELESS COMMUNICATIONS, IEEE ACCESS, Journal of Communications and Information Networks, and IEEE WIRELESS COMMUNICATIONS LETTERS. He served as the Symposium Chair for IEEE GLOBECOM 2019 and IEEE ICCC 2019 and 2020.
	\end{IEEEbiography}
	
	\begin{IEEEbiography}[{\includegraphics[width=1in,height=1.25in,clip,keepaspectratio]{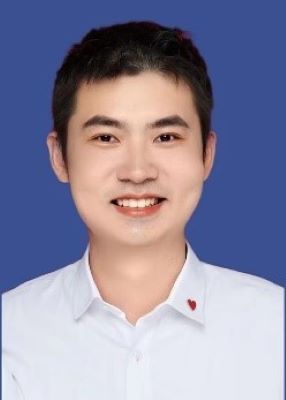}}]{Shulong Zhang}
		received the B.E. degree in Control Science and Engineering from Harbin Institute of Technology, Harbin, China, in 2015, and the Master's degree in Control Science and Engineering from National University of Defense Science and Technology, Changsha, China. He is currently working at SF Technology Co., Ltd. as a system simulation engineer. His current research interests include digital twin and network planning.
	\end{IEEEbiography}
	\vfill
	
	% if you will not have a photo at all:
	%\begin{IEEEbiographynophoto}{author name}
	%Biography text here.
	%\end{IEEEbiographynophoto}
	
	% You can push biographies down or up by placing
	% a \vfill before or after them. The appropriate
	% use of \vfill depends on what kind of text is
	% on the last page and whether or not the columns
	% are being equalized.
	%\vfill
	% Can be used to pull up biographies so that the bottom of the last one
	% is flush with the other column.
	%\enlargethispage{-5in}
	
\end{document}